\documentclass[trackchanges,twocolumn,times]{aastex631}

\usepackage{newtxtext,newtxmath}
\usepackage[T1]{fontenc}
\usepackage{graphicx}	
\usepackage{amsmath}	
\usepackage{float}
\usepackage{bookmark} 
\bookmarksetup{numbered, open,}
\usepackage{enumitem}
\setlist[enumerate]{itemsep=0mm}
\usepackage{hyperref}
\usepackage{subfigure}
\usepackage{xspace}
\usepackage{xcolor} 
\usepackage{soul} 
\usepackage{comment}
\DeclareRobustCommand{\VAN}[3]{#2}
\let\VANthebibliography\thebibliography
\def\thebibliography{\DeclareRobustCommand{\VAN}[3]{##3}\VANthebibliography}

\usepackage{newtxtext,newtxmath}
\usepackage[T1]{fontenc}
\usepackage{orcidlink}
\usepackage{tablefootnote}
\usepackage{todo}
\usepackage{graphicx}
\usepackage[flushleft]{threeparttable}
\usepackage{ulem}
\usepackage{subfigure}
\usepackage{multirow}
\usepackage{array}
\usepackage{hyperref}

\newcolumntype{H}{>{\setbox0=\hbox\bgroup}c<{\egroup}@{}}

\newcommand{\um}{$\mu$m}

\newcommand{\kms}{km~s$^{-1}$}

\newcommand{\Msun}{$M_{\odot}$}

\newcommand{\ms}[1]{{\color{blue} Melissa: {#1}}}

\newcommand{\mic}{\ensuremath{\mu}m\xspace}

\newcommand{\HI}{H~{\sc i}}

\newcommand{\OI}{O~{\sc i}}
\newcommand{\OII}{O~{\sc ii}}

\newcommand{\CII}{C~{\sc ii}}
\newcommand{\CI}{C~{\sc i}}
\newcommand{\NaI}{Na~{\sc i}}
\newcommand{\NI}{N~{\sc i}}
\newcommand{\NII}{N~{\sc ii}}
\newcommand{\NIII}{N~{\sc iii}}
\newcommand{\MgII}{Mg~{\sc ii}}

\newcommand{\SrII}{Sr~{\sc ii}}
\newcommand{\ScII}{Sc~{\sc ii}}
\newcommand{\ScIII}{Sc~{\sc iii}}
\newcommand{\BaII}{Ba~{\sc ii}}

\newcommand{\CaII}{Ca~{\sc ii}}

\newcommand{\FeI}{Fe~{\sc i}}
\newcommand{\FeII}{Fe~{\sc ii}}

\newcommand{\NiII}{Ni~{\sc ii}}

\newcommand{\acko}{SN~2022acko\xspace}

\graphicspath{{./}{Figures/}}

\received{XXX}
\revised{\today}
\accepted{xxx}

\submitjournal{ApJL}

\shorttitle{JWST observations of SN~2022acko}
\shortauthors{Shahbandeh, et al.}

\begin{document}
\title{JWST NIRSpec+MIRI Observations of the nearby Type~IIP supernova~2022acko}

\correspondingauthor{Melissa Shahbandeh}
\email{melissa.shahbandeh@gmail.com}

\author[0000-0002-9301-5302]{M.~Shahbandeh}
\affiliation{Space Telescope Science Institute, 3700 San Martin Drive, Baltimore, MD 21218-2410, USA}

\author[0000-0002-5221-7557]{C.~Ashall}
\affiliation{Department of Physics, Virginia Tech, Blacksburg, VA 24061, USA}

\author[0000-0002-4338-6586]{P.~Hoeflich}
\affiliation{Department of Physics, Florida State University, 77 Chieftan Way, Tallahassee, FL 32306, USA}

\author[0000-0001-5393-1608]{E.~Baron}
\affiliation{Planetary Science Institute, 1700 East Fort Lowell Road, Suite 106,
 Tucson, AZ 85719-2395, USA}
\affiliation{Hamburger Sternwarte, Gojenbergsweg 112, D-21029 Hamburg, Germany}
\affiliation{Dept of Physics \& Astronomy, University of Oklahoma, 440
W. Brooks, Rm 100, Norman, OK USA}

\author[0000-0003-2238-1572]{O.~Fox}
\affiliation{Space Telescope Science Institute, 3700 San Martin Drive, Baltimore, MD 21218-2410, USA}

\author[0000-0001-5888-2542]{T.~Mera}
\affiliation{Department of Physics, Florida State University, 77 Chieftan Way, Tallahassee, FL 32306, USA}

\author[0000-0002-7566-6080]{J.~DerKacy}
\affiliation{Department of Physics, Virginia Tech, Blacksburg, VA 24061, USA}

\author[0000-0002-5571-1833]{M. D.~Stritzinger}
\affiliation{Department of Physics and Astronomy, Aarhus University, Ny Munkegade 120, DK-8000 Aarhus C, Denmark}

\author[0000-0003-4631-1149]{B.~Shappee}
\affiliation{Institute for Astronomy, University of Hawai'i at Manoa, 2680 Woodlawn Dr., Hawai'i, HI 96822, USA }

\author[0000-0002-9402-186X]{D.~Law}
\affiliation{Space Telescope Science Institute, 3700 San Martin Drive, Baltimore, MD 21218-2410, USA}

\author[0000-0002-9288-9235]{J.~Morrison}
\affiliation{Space Telescope Science Institute, 3700 San Martin Drive, Baltimore, MD 21218-2410, USA}

\author[0000-0001-9500-9267]{T.~Pauly}
\affiliation{Space Telescope Science Institute, 3700 San Martin Drive, Baltimore, MD 21218-2410, USA}

\author[0000-0002-2361-7201]{J.~Pierel}
\affiliation{Space Telescope Science Institute, 3700 San Martin Drive, Baltimore, MD 21218-2410, USA}

\author[0000-0001-7186-105X]{K.~Medler}
\affiliation{Astrophysics Research Institute, Liverpool John Moores University, UK}

\author[0000-0003-0123-0062]{J.~Andrews}
\affiliation{Gemini Observatory/NSF’s NOIRLab, 670 North A`ohoku Place, Hilo, HI 96720-2700, USA}

\author[0000-0003-1637-9679]{D.~Baade}
\affiliation{European Organization for Astronomical Research in the Southern Hemisphere (ESO), Karl-Schwarzschild-Str. 2, 85748 Garching b. M\"unchen, Germany}

\author[0000-0002-4924-444X]{A.~Bostroem}
\altaffiliation{LSSTC Catalyst Fellow}
\affiliation{Steward Observatory, University of Arizona, 933 North Cherry Avenue, Tucson, AZ 85721-0065, USA}

\author[0000-0001-6272-5507]{P.~Brown}
\affiliation{George P. and Cynthia Woods Mitchell Institute for Fundamental Physics and Astronomy, Texas A\&M University, Department of Physics and Astronomy, College Station, TX 77843, USA}

\author[0000-0003-4625-6629]{C.~Burns}
\affiliation{Observatories of the Carnegie Institution for Science, 813 Santa Barbara Street, Pasadena, CA 91101, USA}

\author[0000-0002-5380-0816]{A.~Burrow}
\affiliation{Homer L. Dodge Department of Physics and Astronomy, University of Oklahoma, 440 W. Brooks, Rm 100, Norman, OK 73019-2061, USA}

\author[0000-0001-7101-9831]{A.~Cikota}
\affiliation{Gemini Observatory/NSF's NOIRLab, Casilla 603, La Serena, Chile}

\author{D.~Cross}
\affiliation{Institute of Space Sciences (ICE, CSIC), Campus UAB, Carrer de Can Magrans, s/n, E-08193 Barcelona, Spain}

\author[0000-0003-4631-1149]{S.~Davis}
\affiliation{Private Astronomer}

\author[0000-0001-6069-1139]{T.~de~Jaeger}
\affiliation{LPNHE, (CNRS/IN2P3, Sorbonne Universit\'{e}, Universit\'{e} Paris Cit\'{e}), Laboratoire de Physique Nucl\'{e}aire et de Hautes \'{E}nergies, 75005, Paris, France}

\author[0000-0003-3429-7845]{A.~Do}
\affiliation{Institute for Astronomy, University of Hawai'i at Manoa, 2680 Woodlawn Dr., Hawai'i, HI 96822, USA }

\author[0000-0002-7937-6371]{Y.~Dong}
\affiliation{Department of Physics, University of California, 1 Shields Avenue, Davis, CA 95616-5270, USA}

\author[0000-0002-3827-4731]{I.~Dominguez}
\affiliation{Universidad de Granada, 18071, Granada, Spain}

\author[0000-0002-1296-6887]{L.~Galbany}
\affiliation{Institute of Space Sciences (ICE, CSIC), Campus UAB, Carrer de Can Magrans, s/n, E-08193 Barcelona, Spain}
\affiliation{Institut d’Estudis Espacials de Catalunya (IEEC), E-08034 Barcelona, Spain}

\author[0000-0003-0549-3281]{D.~Janzen}
\affiliation{Department of Physics and Engineering Physics, University of Saskatchewan, 116 Science Place, Saskatoon, SK S7N 5E2, Canada}

\author[0000-0001-5754-4007]{J.~Jencson}
\affiliation{Department of Physics and Astronomy, The Johns Hopkins University, 3400 North Charles Street, Baltimore, MD 21218, USA}

\author[0000-0003-2744-4755]{E.~Hoang}
\affiliation{Department of Physics and Astronomy, University of California, Davis, 1 Shields Avenue, Davis, CA 95616-5270, USA}

\author[0000-0003-1039-2928]{E.~Hsiao}
\affiliation{Department of Physics, Florida State University, 77 Chieftan Way, Tallahassee, FL 32306, USA}

\author[0000-0001-6209-838X]{E.~Karamehmetoglu}
\affiliation{Department of Physics and Astronomy, Aarhus University, Ny Munkegade 120, DK-8000 Aarhus C, Denmark}

\author[0009-0005-0311-0058]{B.~Khaghani}
\affiliation{Department of Physics, Virginia Tech, Blacksburg, VA 24061, USA}

\author[0000-0002-6650-694X]{K.~Krisciunas}
\affiliation{George P. and Cynthia Woods Mitchell Institute for Fundamental Physics and Astronomy, Texas A\&M University, Department of Physics and Astronomy, College Station, TX 77843, USA}

\author[0000-0001-8367-7591]{S.~Kumar}
\affiliation{Department of Physics, Florida State University, 77 Chieftan Way, Tallahassee, FL 32306, USA}

\author[0000-0002-3900-1452]{J.~Lu}
\affiliation{Department of Physics, Florida State University, 77 Chieftan Way, Tallahassee, FL 32306, USA}

\author[0000-0001-9589-3793]{M.~Lundquist}
\affiliation{W.~M.~Keck Observatory, 65-1120 M\=amalahoa Highway, Kamuela, HI 96743-8431, USA}

\author[0000-0003-0733-7215]{J.~Maund}
\affiliation{Department of Physics and Astronomy, University of Sheffield, Hicks Building, Hounsfield Road, Sheffield S3 7RH, U.K.}

\author[0000-0001-6876-8284]{P.~Mazzali}
\affiliation{Astrophysics Research Institute, Liverpool John Moores University, UK}
\affiliation{Max-Planck Institute for Astrophysics, Garching, Germany}

\author[0000-0003-2535-3091]{N.~Morrell}
\affiliation{Las Campanas Observatory, Carnegie Observatories, Casilla 601, La Serena, Chile}
	
\author[0000-0002-0537-3573]{F.~Patat}
\affiliation{European Organization for Astronomical Research in the Southern Hemisphere (ESO), Karl-Schwarzschild-Str. 2, 85748 Garching b. M\"unchen, Germany}

\author[0000-0002-0744-0047]{J.~Pearson}
\affiliation{Steward Observatory, University of Arizona, 933 North Cherry Avenue, Tucson, AZ 85721-0065, USA}

\author[0000-0002-7305-8321]{C.~ Pfeffer}
\affiliation{Department of Physics, Virginia Tech, Blacksburg, VA 24061, USA}

\author[0000-0003-2734-0796]{M.~Phillips}
\affiliation{Las Campanas Observatory, Carnegie Observatories, Casilla 601, La Serena, Chile}

\author{A.~Rest}
\affiliation{Space Telescope Science Institute, 3700 San Martin Drive, Baltimore, MD 21218-2410, USA}

\author[0000-0002-7015-3446]{N.~Retamal}
\affiliation{Department of Physics, University of California, 1 Shields Avenue, Davis, CA 95616-5270, USA}

\author[0000-0001-5570-6666]{S.~Stangl}
\affiliation{Homer L. Dodge Department of Physics and Astronomy, University of Oklahoma, 440 W. Brooks, Rm 100, Norman, OK 73019-2061, USA}

\author[0000-0002-4022-1874]{M.~Shrestha}
\affiliation{Steward Observatory, University of Arizona, 933 North Cherry Avenue, Tucson, AZ 85721-0065, USA}

\author[0000-0003-0763-6004]{C.~Stevens}
\affiliation{Department of Physics, Virginia Tech, Blacksburg, VA 24061, USA}

\author[0000-0002-8102-181X]{N.~Suntzeff}
\affiliation{George P. and Cynthia Woods Mitchell Institute for Fundamental Physics and Astronomy, Texas A\&M University, Department of Physics and Astronomy, College Station, TX 77843, USA}

\author[0000-0002-0036-9292]{C.~Telesco}
\affiliation{Department of Astronomy, University of Florida, Gainesville, FL 32611 USA}

\author[0000-0002-2471-8442]{M.~Tucker}
\altaffiliation{CCAPP Fellow}
\affiliation{Center for Cosmology and AstroParticle Physics, The Ohio State University, 191 W. Woodruff Ave., Columbus, OH 43210, USA}

\author[0000-0001-7092-9374]{L.~Wang}
\affiliation{Department of Physics and Astronomy, Texas A\&M University, College Station, TX 77843, USA}

\author[0000-0002-6535-8500]{Y.~Yang}
\affiliation{Department of Astronomy, University of California, Berkeley, CA 94720-3411, USA}
\altaffiliation{Bengier-Winslow-Robertson Postdoctoral Fellow}

\author{Y.-Z.~Cai}
\affiliation{Yunnan Observatories, Chinese Academy of Sciences, Kunming
650216, PR China}
\affiliation{Key Laboratory for the Structure and Evolution of Celestial Objects, Chinese Academy of Sciences, Kunming 650216, PR China}
\affiliation{International Centre of Supernovae, Yunnan Key Laboratory, Kunming 650216, P.R. China}

\author[0000-0002-9830-3880]{Y.~Camacho-Neves}
\affiliation{Department of Physics and Astronomy, Rutgers, the State University of New Jersey,\\136 Frelinghuysen Road, Piscataway, NJ 08854-8019, USA}

\author[0000-0002-1381-9125]{N.~Elias-Rosa}
\affiliation{INAF - Osservatorio Astronomico di Padova, Vicolo
dell’Osservatorio 5, 35122, Padova, Italy}

\author[0000-0002-2445-5275]{R.~Foley}
\affiliation{Department of Astronomy and Astrophysics, University of California, Santa Cruz, CA 95064-1077, USA}

\author[0000-0001-8738-6011]{S.~Jha}
\affiliation{Department of Physics and Astronomy, Rutgers, the State University of New Jersey,\\136 Frelinghuysen Road, Piscataway, NJ 08854-8019, USA}

\author[0000-0003-3108-1328]{L.~Kwok}
\affiliation{Department of Physics and Astronomy, Rutgers, the State University of New Jersey,\\136 Frelinghuysen Road, Piscataway, NJ 08854-8019, USA}

\author[0000-0003-2037-4619]{C.~Larison}
\affiliation{Department of Physics and Astronomy, Rutgers, the State University of New Jersey,\\136 Frelinghuysen Road, Piscataway, NJ 08854-8019, USA}

\author{N.~LeBaron}
\affiliation{Department of Astronomy, University of California, Berkeley, CA 94720-3411, USA}

\author[0000-0001-5221-0243]{S.~Moran}
\affiliation{Department of Physics and Astronomy, University of Turku, Vesilinnantie 5, FI-20500, Finland}

\author{J.~Rho}
\affiliation{SETI Institute, 339 N. Bernardo Ave., Ste. 200, Mountain View, CA 94043, USA}
\affil{Department of Physics and Astronomy, Seoul National University, Gwanak-ro 1, Gwanak-gu, Seoul, 08826, South Korea}

\author[0000-0003-1450-0869]{I.~Salmaso}
\affiliation{Dipartimento di Fisica e Astronomia ``G. Galilei'', Università degli Studi di Padova, Vicolo dell'Osservatorio 3, 35122, Padova, Italy}
\affiliation{INAF - Osservatorio Astronomico di Padova, Vicolo dell'Osservatorio 5, 35122, Padova, Italy}

\author{J.~Schmidt}
\affiliation{Private Astronomer}

\author[0000-0002-1481-4676]{S.~Tinyanont}
\affiliation{National Astronomical Research Institute of Thailand, 260 Moo 4, Donkaew, Maerim, Chiang Mai 50180, Thailand}

\begin{abstract}
We present \textit{JWST} spectral and photometric observations of the Type~IIP supernova (SN) 2022acko at $\sim$50~days past explosions. These data are the first \textit{JWST} spectral observations of a core-collapse SN. 
We identify $\sim30$ different \HI\ features, other features associated with products produced from the CNO cycle, and s-process elements such as \ScII,
\BaII. By combining the JWST spectra with ground-based
optical and NIR  spectra, we construct a full Spectral Energy
Distribution from 0.4 to 25~$\micron$ and find that the \textit{JWST}
spectra are fully consistent with the simultaneous \textit{JWST} photometry.  The data lack signatures of  $CO$ formation and we estimate a limit on the $CO$ mass of $< 10^{-8} M_\odot$. 
We demonstrate how the $CO$ fundamental band limits can be used to probe underlying physics during the stellar evolution, explosion, and the environment. 
The observations indicate little mixing between the H envelope and C/O core in the ejecta and show no evidence for dust.
The data presented here set a critical baseline for future \textit{JWST} observations, where possible molecular and dust formation may be seen. 
\end{abstract}

\keywords{supernovae: general - supernovae: individual (SN~2022acko), JWST}

\section{Introduction} \label{sec:intro}

The origin of the substantial quantities of dust inferred from submillimeter and millimeter observations of high-redshift ($z$) quasars is an unsolved astrophysical question 
\citep[e.g.,][]{Bertoldi2003,Priddey2003,Li2020}.  In the local Universe, most dust is believed to 
originate during the evolutionary phase of low- and intermediate-mass stars when they reach the asymptotic giant branch on the Hertzsprung-Russell diagram \citep[e.g.,][]{Gonzo2003,Ferrarotti2006}, However, at higher redshifts (i.e., $z > 6.3$), alternative processes occurring on shorter time scales are more likely at play. 

For more than half a century, core-collapse supernovae (CCSNe) have been proposed as a potential source of dust \citep{Cernuschi_1967, Hoyle_1970}. In particular, hydrogen-rich CCSNe known as Type~II plateau supernovae (hereafter, SNe~IIP), are linked to the demise of massive stars born with initial masses ranging between 8 to 25~\Msun, and with lifetimes of less than 50~Myrs. SNe~IIP have played a significant role in the evolution of the Universe, contributing significantly to the production of heavy elements, and the genesis of dust, both locally and at cosmologically significant redshifts \citep[e.g.,][]{Maiolino_2004,Schneider_2004, Dwek_2007, Nozawa_2008, Gall_2011_a}. 
Models of the expanding SN ejecta have demonstrated their ability to condense sufficient quantities (0.1--1 $M_{\sun}$)  of dust  \citep[e.g.,][]{Todini_2001, Nozawa_2003, Cherchneff_2009, Sarangi_2015, Sluder_2018, Sarangi_2018}.

Only a handful of SNe~IIP have been documented to form dust,  and until recently, these cases produced 2--3 orders of magnitude less dust than predicted by modeling.
The prerequisite for the genesis of dust is the creation of molecules, which serve both as a coolant and as a nucleation site.  In SNe~IIP, the order of the formation of silicon-monoxide ($SiO$) versus carbon-monoxide ($CO$) can be dependent upon the initial zero-age main sequence (ZAMS) mass of the progenitor stars since it leads to differing amounts of C, O, and Si in the stars at the time of their demise \citep{Woosley_2002}.  

The majority of previous studies on molecular formation in CCSNe have concentrated on searching for the ($\approx2.3~\micron$) $CO$ first overtone feature.
Evidence of molecular formation was first found in data of SN~1987A  \citep[e.g.,][]{Catchpole1988,Meikle1989,Wooden1993}, and to date has also been identified in around a dozen SNe~IIP \citep[e.g.,][]{Aitken88,Kotak2005,2007ApJ...665..608M,2009ApJ...704..306K,Fox10,Tinyanont19,Szalai19,Davis_2019}.

The ability to observe the $CO$ and $SiO$ fundamental bands  began to improve with the advent of the {\it Spitzer Space Telescope} ({\it SST}), as the longer spectral wavelength range was particularily well-suited for studying molecular formation in CCSNe.
{\it SST} spectra of SN~2005af showed indications of  $SiO$ formation \citep{Kotak_etal_2006_05af}, 
while observations of SN~2004dj contained the red part of the $CO$ fundamental band \citep{Kotak_etal_2005_04dj}. Given the nature of the data, these initial assessments of molecules in a handful of SNe left open the 
question of whether we see pre-existing or freshly synthesized dust \citep{Gerardy_etal_2000}.

The {\it James Webb Space Telescope (JWST)} has revolutionized the communities ability to accurately study molecule and dust formation in various types of SNe. 
Recently, \citet{Shahbandeh_2023} used {\it JWST} Mid-Infrared Instrument (MIRI; \citealt{Bouchet_2015, Ressler_2015, Rieke_2015, Rieke_2022})  observations of the Type~IIP SN~2004et to infer a dust mass lower limit of $0.014$~\Msun.
 This exceeds estimates  estimated from  Mid-infrared 
 (MIR) observations of all other observed SNe IIP \citep[see][]{Shahbandeh_2023}
 , and highlights the potential of SNe~IIP being a major contributor to the dust content of the Universe. 

While deep imaging is possible, {\it JWST} spectroscopy offers additional insights into the explosion physics of  CCSNe.  Near-infrared (NIR) and MIR spectroscopic time series covering the duration of a CCSN's lifetime can provide information on the formation of molecules like $CO$ and $SiO$. To fully understand how SNe~IIP produces molecules and dust, we must follow them beginning soon after exploding and over the duration of many decades.  
Where early data allows us to set a baseline for any pre-existing molecular and dust formation.


We present Near-Infrared Spectrograph (NIRSpec; \citealt{Jakobsen_2022}) and MIRI observations of the Type~IIP SN~2022acko, as part of {\it JWST} GO-2122 (PI: C. Ashall), which aims to follow the evolution of a single SN~IIP from the early H recombination driven plateau phase thru late times when $CO$ and $SiO$ molecules form. 
Here, we present the first NIRSpec+MIRI spectral observation from this program, the first CCSN spectrum obtained with {\it JWST}. 
We also present near-simultaneous optical and NIR ground-based spectral observations of SN~2022acko, as well as serendipitous multi-band imaging obtained by the Physics at High Angular Resolution of Nearby GalaxieS (PHANGS) \textit{JWST} Cycle 1 Treasury Program \citep{Lee23}. 

Our goals are to \textit{i}) measure the level of mixing within the ejecta by searching for $CO$ formation, trace the elements of the CNO-cycle, and identify possible s-process elements, \textit{ii}) set limits for pre-existing molecules and dust in the interstellar medium (ISM) and circumstellar medium (CSM), and \textit{iii}) demonstrate the importance of fundamental bands of molecules, obtain values and tight upper limits for the nearby environment, for mixing processes during the stellar evolution, and for the physics of the explosion. 
This is a key to being able to separate pre-existing molecules and dust from newly formed products, and to evaluate the SN~IIP as dust-producers in the early Universe. 


SN~2022acko was discovered on 2022 December 06.2 UT  (i.e., MJD-59919.2)   by the Distance Less Than 40 Mpc Survey (DLT40; \citealt{DLT40}), and a spectrum obtained 12 hours with the Lijiang 2.4-m telescope showed it to be a young SN~IIP \citep{Li22}. 
With J2000 coordinates R.A. $= 03^{h}19^{m}38^{s}.99$ and 
Decl. $=  -19^\circ23'42\farcs68$, the location of SN~2022acko is $0.0181^\circ$ North-West from the center of NCG~1300 \citep{1991rc3..book.....D}. 
According to the NASA extragalactic database (NED) the heliocentric redshift ($z$) to the host-galaxy of \acko\ is $z = 0.00526$. In the following, we adopt the tip of red giant branch  (TRGB) distance measurement from the PHANGS team of $18.99 \pm 2.85$~Mpc \citep{2021MNRAS.501.3621A}.

Its proximity as well as the location of \acko\ outside the core of its host made it an ideal target for our  {\it JWST}-GO-2122  ToO program  (PI: C. Ashall, DOI:\href{https://archive.stsci.edu/doi/resolve/resolve.html?doi=10.17909/ea3g-5z06}{10.17909/ea3g-5z06}).
Ultraviolet and optical observations of  SN~2022acko  indicate it  is a low-luminosity SN~IIP \citep[see][]{Bostroem23}.  
For consistency, throughout this work, we adopt the same basic parameters for \acko\ as inferred by \citeauthor{Bostroem23} (see Table~\ref{tab:22ackodetails}),  including their preferred reddening values and estimated explosion epoch which they demonstrated likely occurred within 24 hours prior to discovery.

\begin{deluxetable}{lcc}
  \tablecaption{SN~2022acko details. \label{tab:22ackodetails}} 
  \tablehead{\colhead{Parameter} & \colhead{Value} & \colhead{Reference}}
  \startdata
  RA &03:19:38.99&\citet{2022TNSTR3543....1L}\\
Dec & $-$19:23:42.68&\citet{2022TNSTR3543....1L}\\
  Redshift & 0.00526&\citet{Bostroem23}\\
  Host galaxy & NGC~1300&\citet{2022TNSTR3543....1L}\\
$E(B-V)_{MW}$  & 0.026 $\pm$0.001 mag&\citet{Bostroem23}\\
$E(B-V)_{host}$& $0.04 \pm 0.01$ mag&\citet{Bostroem23}\\
Distance [Mpc]& $19.0\pm2.9$ &\citet{2021MNRAS.501.3621A}\\ 
Explosion epoch ($t_{exp}$; MJD)\tablenotemark{a}& $59918.17\pm0.4$ &\citet{Bostroem23}\\ 
    \enddata
\end{deluxetable}

The outline of this paper is as follows.  Our observations of \acko\ are presented in Section~\ref{sec:obs}, which is then followed by the identification of key spectral features and data comparisons in Section~\ref{sec:IDs}.
Next in  Section~\ref{sec:Hvel}, H line velocities are measured, 
after which the full optical-to-MIR spectral energy distribution (SED) constructed in Section~\ref{sec:SED}. 
Limits on pre-existing molecular and dust formation are determined in Section~\ref{sec:COformation}, and then in Section~\ref{sec:1987Acomparison}, \acko\ is compared with SN~1987A in order to understand the amount of mixing in the ejecta.
Finally, a summary of the results and our conclusions are given in Section~\ref{sec:conclusions}.

\section{Observations}

\label{sec:obs}

\subsection{NIRCam and MIRI Imaging}

PHANGS obtained imaging of NGC~1300 with NIRCam and MIRI on 59969.74~MJD (51.57~days post-explosion), which is within  $\sim 1$~day of our programme's NIRSpec observations of \acko (see below). 
This data was acquired in four NIR filters (i.e., \textit{F200W}, \textit{F300M}, and \textit{F335M}, \textit{F360M}), as well as in four MIR filters (i.e., \textit{F770W}, \textit{F1000W}, \textit{F1130W}, and \textit{F2100W}) providing a measure of SN~2022acko's Spectral Energy Distribution (SED) between $\sim$1.7--21.0~\um. 
Calibrated level 2 data products were downloaded from the Mikulski Archive for Space Telescopes (MAST), from which point-spread function (PSF) photometry was computed. 

\begin{figure*}
    \centering
    \includegraphics[width=0.95\textwidth]{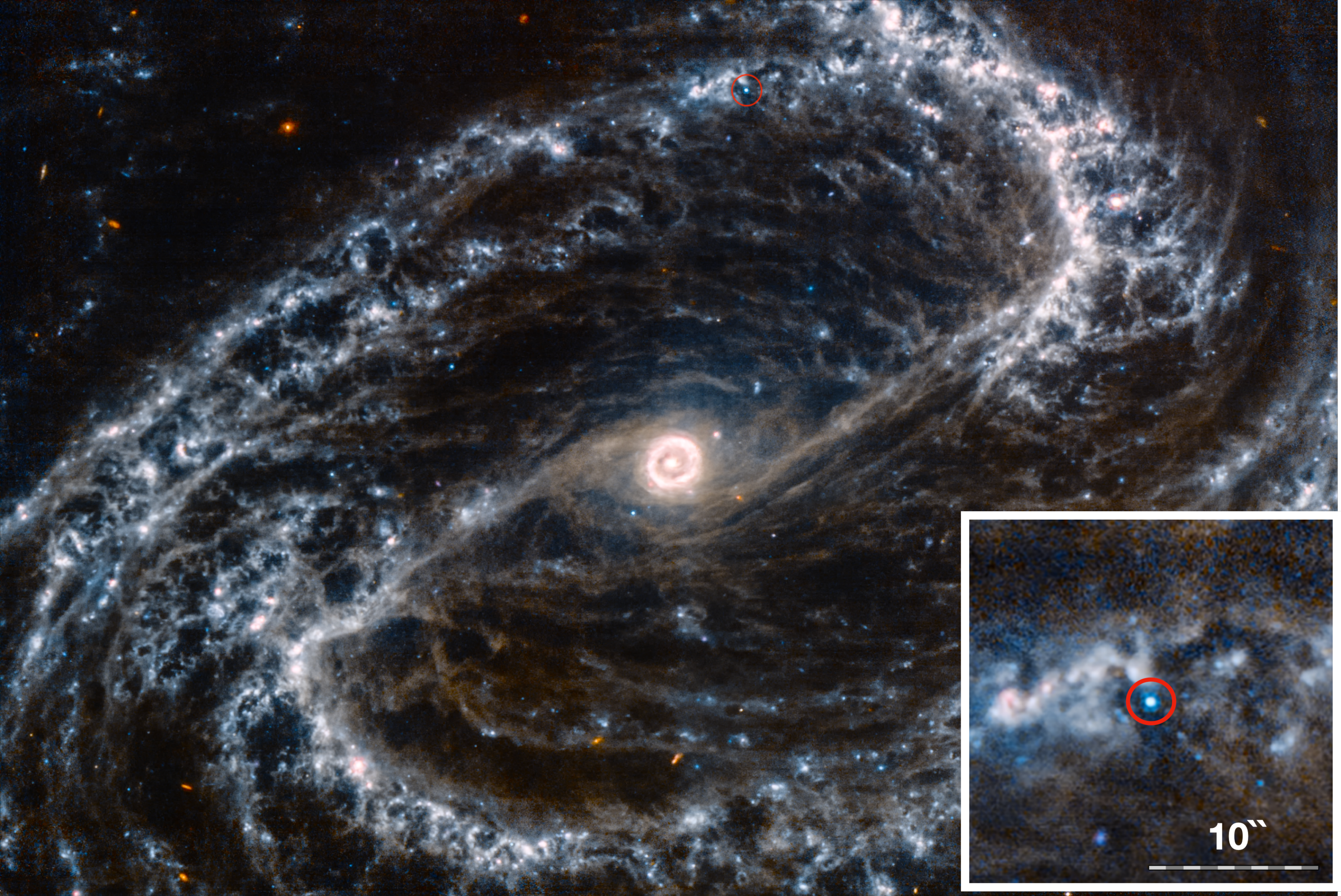}
    \caption{A color image of NGC~1300 constructed from three \textit{JWST} MIRI images. The color code of the images are is followed, Red: \textit{F2100W}, Orange: \textit{F1130W}, Cyan: \textit{F770W}. SN~2022acko is located in the top of the image in the red circle. A 10'' by 10'' inset around the location of the SN is seen in the bottom right. The SN is located in an area of star formation. The full resolution of this image can be found \href{https://www.flickr.com/photos/geckzilla/52701175433/}{here.} }
    \label{fig:colorimage}
\end{figure*}

\begin{figure}
    \centering
    \includegraphics[width=0.55\textwidth]{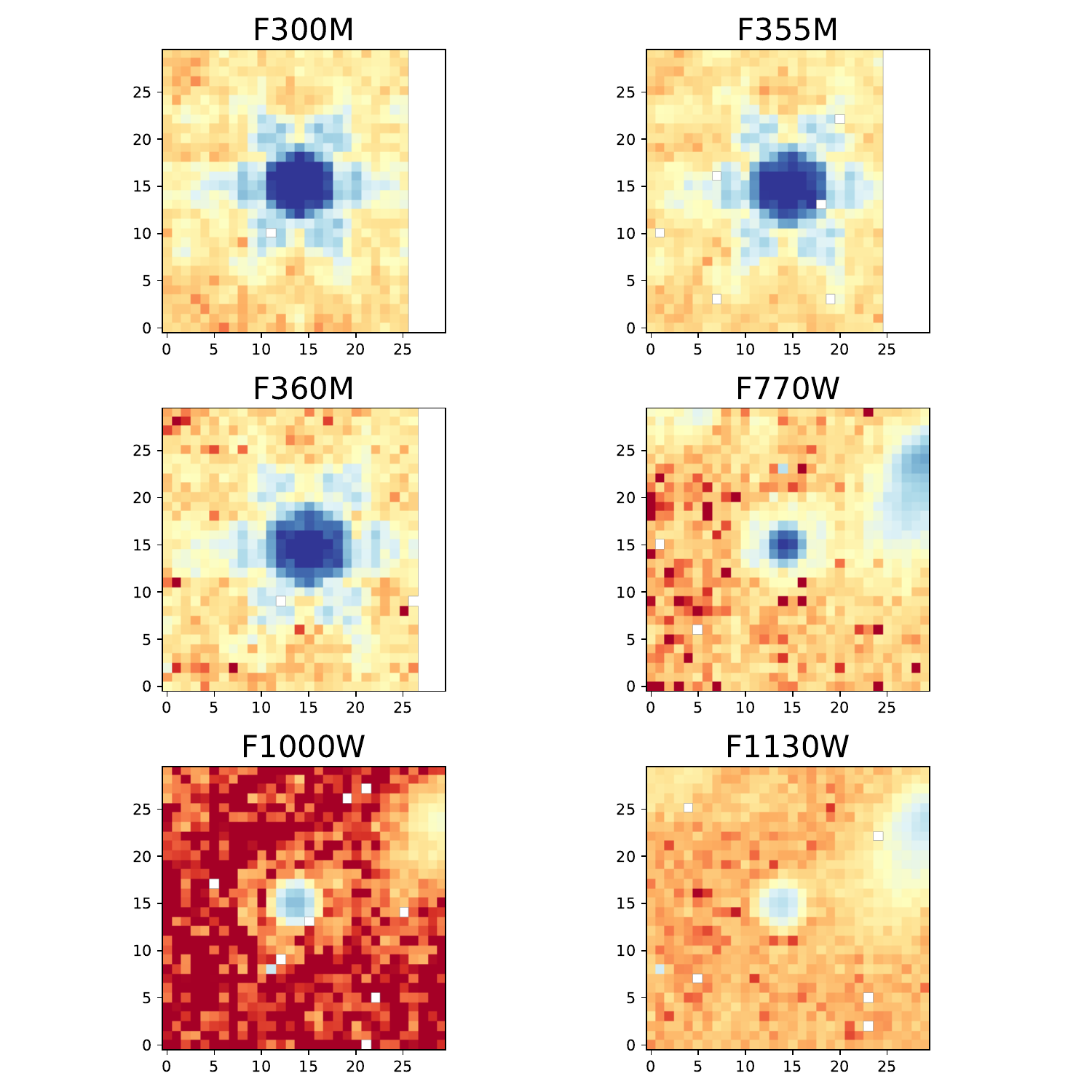}
    \caption{NIRCAM \textit{F300M}, \textit{F355M}, \textit{F360M} 30 pixel $\times$ 30 pixel stamps of SN~2022acko, and MIRI \textit{F770W}, \textit{F1000W} and \textit{F1130W} 30 pixel $\times$ 30 pixel stamps of SN~2022acko. We do not show the \textit{F2100W} filter as the SN is not visible. }
    \label{fig:stamps}
\end{figure}

A color-stacked image constructed from MIRI images is presented in  Fig.~\ref{fig:colorimage}. 
The position of SN~2022acko  at the top of the image is indicated with a red circle while a  $10\times10$ arcsec zoom of this region is provided in the bottom-right inset. 
\acko\ is located in a region of star formation. 
Figure~\ref{fig:stamps} shows the SN in a 30 by 30 pixel stamp
in the $F300M$, $F355M$, $F360M$, $F770W$, $F1000W$ and $F1130W$ filters.  
The SN is clearly visible in each of these bands.


PSF photometry was computed using a coustom-made phython notebook making use of the  \textit{WebbPSF} photometry package \citep{Perrin14}.\footnote{\url{https://github.com/orifox/psf_phot/blob/main/space_phot/MIRI/miri_1028.ipynb}}
Various PSF sizes were used to ensure that the full variation within the data was encapsulated.
The final magnitude was taken as the average flux across all fits, which have a residual less than 15\%.
The fluxes of all four dithers of each filter were then averaged, and the error in the flux was obtained using the standard deviation of the fluxes. A log of the photometric observations and the flux at these epochs, can be found in Table~\ref{tab:JWST_phot}. 


\begin{table}
\centering
\caption{ Log of {\it JWST} photometric observations. The average time of photometric observation is 51.6~d after the inferred time of the explosion.\label{tab:JWST_phot}  }
\begin{tabular}{ l c c c h}
\hline
\hline
Filters&Obs. date&Exp. time&Mag & Flux \\
& [MJD] & [s] & [mag] &[mJy]\\
\hline
\textit{F200W}\tablenotemark{a} &59969.82&215&17.00 $\pm$ 0.09\tablenotemark{c}&0.57\ms{$\pm$XX}\\
\textit{F300M}\tablenotemark{a} &59969.79&387&17.53 $\pm$ 0.03 & 0.35\ms{$\pm$XX} \\
\textit{F335M}\tablenotemark{a} &59969.80&387&17.60 $\pm$ 0.05 &0.33\ms{$\pm$XX}\\
\textit{F360M}\tablenotemark{a} &59969.77&215\tablenotemark{d}&17.62 $\pm$ 0.01 &0.32\ms{$\pm$XX}\\
\textit{F770W}\tablenotemark{b} & 59969.65&88.8&18.63 $\pm$ 0.01&0.13\ms{$\pm$XX}\\
\textit{F1000W}\tablenotemark{b} &59969.66&122&19.19 $\pm$ 0.01&0.08\ms{$\pm$XX}\\
\textit{F1130W}\tablenotemark{b} &59969.66&311&19.35 $\pm$ 0.01&0.07\ms{$\pm$XX}\\
\textit{F2100W}\tablenotemark{b} &59969.68&322&$\leq$21.43 $\pm$ $\cdots$&0.01\ms{$\pm$XX}\\
\hline
\end{tabular}
\tablecomments{~\tablenotemark{a}NIRCAM data, ~\tablenotemark{b}MIRI data. All photometric observations were performed with a 4-point dither. ~\tablenotemark{c} The F200W images were saturated in all exposures, therefore the photometry is highly uncertain.  All exposure times are reported from four dithers unless otherwise stated.  }
\end{table}


\subsection{NIRSpec}
We obtained a NIRSpec spectrum (R$\sim$~1000) of \acko\ from $\sim$1.7-5.2~$\micron$. 
The spectrum was acquired using the S400A1 slit, the G235M and G395M gratings, with the F170P and F290LP filters, respectively. 
The instrument setup had one exposure per grating and a total exposure time of 467.65~s.  
The details of the observational setup can be seen in Table \ref{tab:obs}. 
The data were taken on MJD 59968.4 at 50.2~d past explosion.
The observation consisted of a three-point dither pattern. 
The background-subtracted level-two spectra were combined into a single calibrated spectrum for each filter.
The spectra were processed with the {\it JWST} Calibration Pipeline version 1.8.2, using the Calibration Reference Data System version 11.16.14 \citep{Bushouse2023_JWSTpipeline}.

\begin{deluxetable}{lccc}
  \tablecaption{SN~2022acko spectral log.  \label{tab:obs}} 
  \tablehead{\colhead{Parameter} & \colhead{Value}& \colhead{Value}& \colhead{Value}}
  \startdata
    \multicolumn{4}{c}{MIRI/MRS Spectra}  \\
        \hline
    Grating& Short &Medium & Long\\
    Groups per Integration& 36&36&36 \\
    Integrations per Exp. &1&1&1\\
    Exposures per Dither &1&1&1\\
    Total Dithers & 4 & 4& 4\\
    Exp Time [s] & 3440.148& 3440.148 & 3440.148 \\
    Resolution\tablenotemark{a} &  \multicolumn{3}{c}{$\sim$2,700}\\ 
    Epoch\tablenotemark{a} [days] &  \multicolumn{3}{c}{55.57}\\ 
    $T_\text{obs}$ [MJD] & \multicolumn{3}{c}{59973.74} \\  
    \hline
     \multicolumn{4}{c}{NIRSpec Acquisition Image}  \\
    \hline
    Filter & CLEAR& $\cdots$& $\cdots$ \\
    Exp Time [s] & 0.08& $\cdots$& $\cdots$ \\
    Readout Pattern & NRSRAPID & $\cdots$& $\cdots$ \\
    \hline
    \multicolumn{4}{c}{NIRSpec Spectra}  \\
    \hline
    Slit &S400A1&S400A1 & $\cdots$\\
    Filter & F170P &F290LP & $\cdots$\\
    Groups per Integration & 5 &7 & $\cdots$\\
    Integrations per Exp. & 2 &2 & $\cdots$\\
    Exposures per Dither & 1 &2 & $\cdots$\\
    Total Dithers & 3 & 3 & $\cdots$\\
    Exp Time [s] & 196.43& 271.22 & $\cdots$ \\
    Epoch\tablenotemark{a} [days] & 50.18  & 50.19 &$\cdots$\\ 
    $T_\text{obs}$ [MJD] &59968.35&59968.36&$\cdots$ \\
      \hline
     \multicolumn{4}{c} {NIR Spectra/Keck-NIRES}\\
        \hline
         Exp time [s] &1200&$\cdots$& $\cdots$\\
        $T_\text{obs}$  [MJD] &59951.95&$\cdots$& $\cdots$\\
        Epoch\tablenotemark{a} [days]&33.78&$\cdots$& $\cdots$\\
        Total Dithers & 4&$\cdots$&$\cdots$\\
        Silt width &0.5''&$\cdots$& $\cdots$\\
              \hline
        \multicolumn{4}{c} {Optical Spectra/NOT-ALFOSC}\\
        \hline
        Exp time [s] &  1800&$\cdots$& $\cdots$\\
        $T_\text{obs}$  [MJD] &59978.86&$\cdots$& $\cdots$\\
        Epoch\tablenotemark{a} [days]&60.69&$\cdots$& $\cdots$\\
        Grism&Grism\_\#4&$\cdots$& $\cdots$\\
        Silt width &1.0''&$\cdots$& $\cdots$\\
              \hline
  \enddata
  \tablecomments{~\tablenotemark{a}~Rest frame days relative to 
  time of explosion, MJD $=59918.17$ \citep{Bostroem23}.}
\end{deluxetable}

\begin{figure*}
    \centering
    \includegraphics[width=0.99\textwidth]{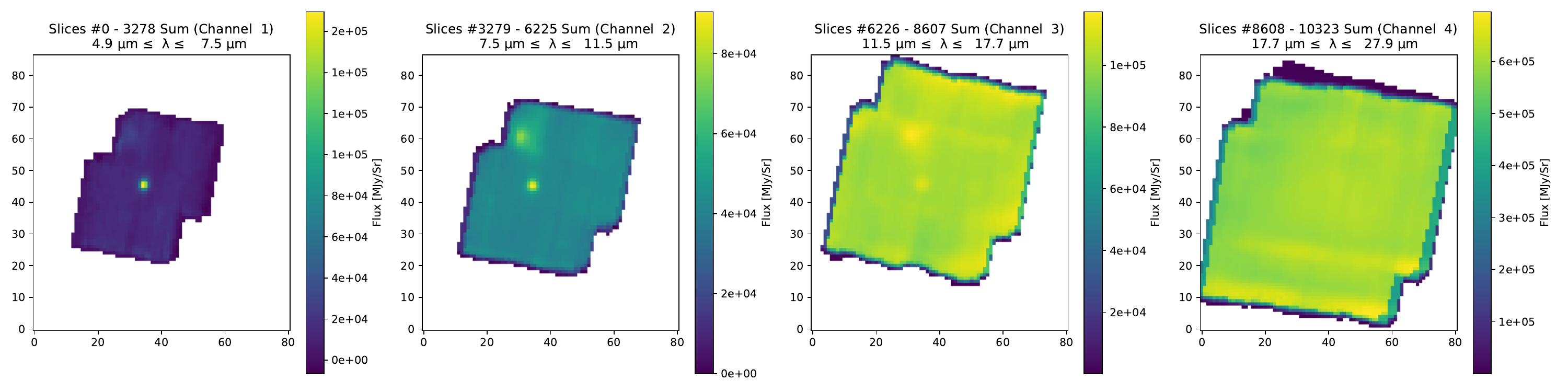} 
     \includegraphics[width=0.99\textwidth]{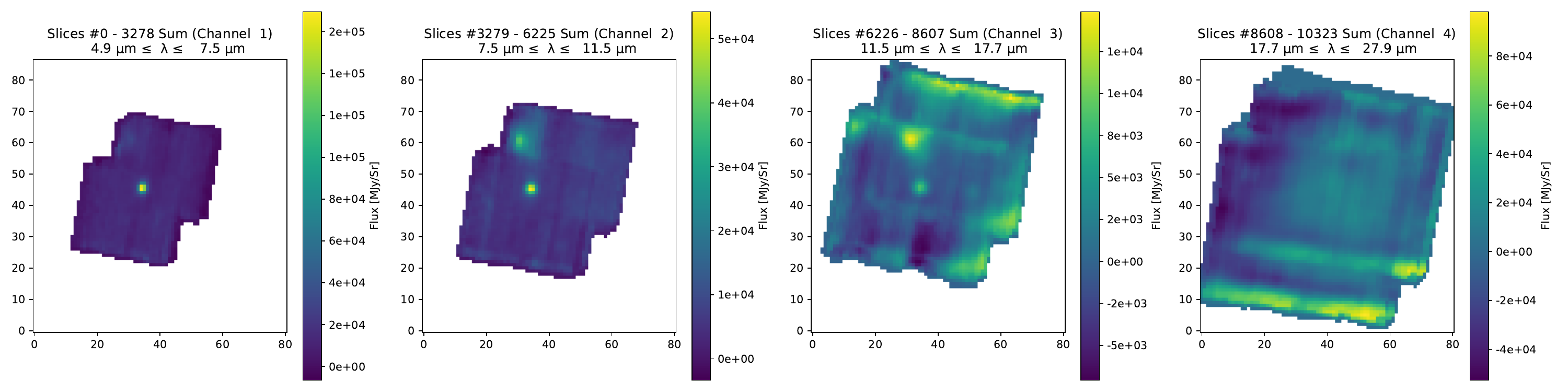}
    \caption{\textit{Top panel:} MIRI/MRS cube of SN~2022acko before background subtraction divided seperate by channel. Each channel is a collapsed sum of all its slices.
    \textit{Bottom panel:} MIRI/MRS cubes of SN~2022acko after background subtraction.}
    \label{fig:cubes}
\end{figure*}

\begin{figure*}
    \centering
    \includegraphics[width=0.85\textwidth]{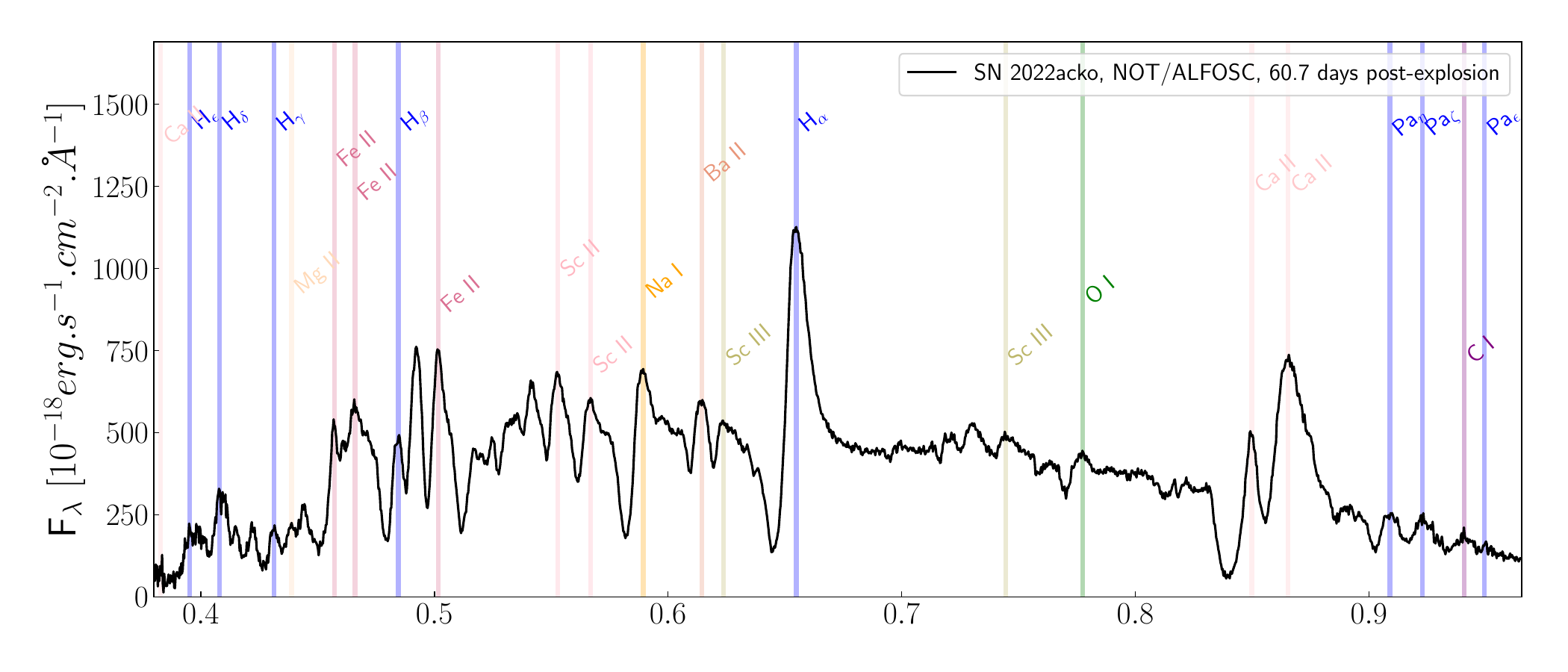}
     \includegraphics[width=0.85\textwidth]{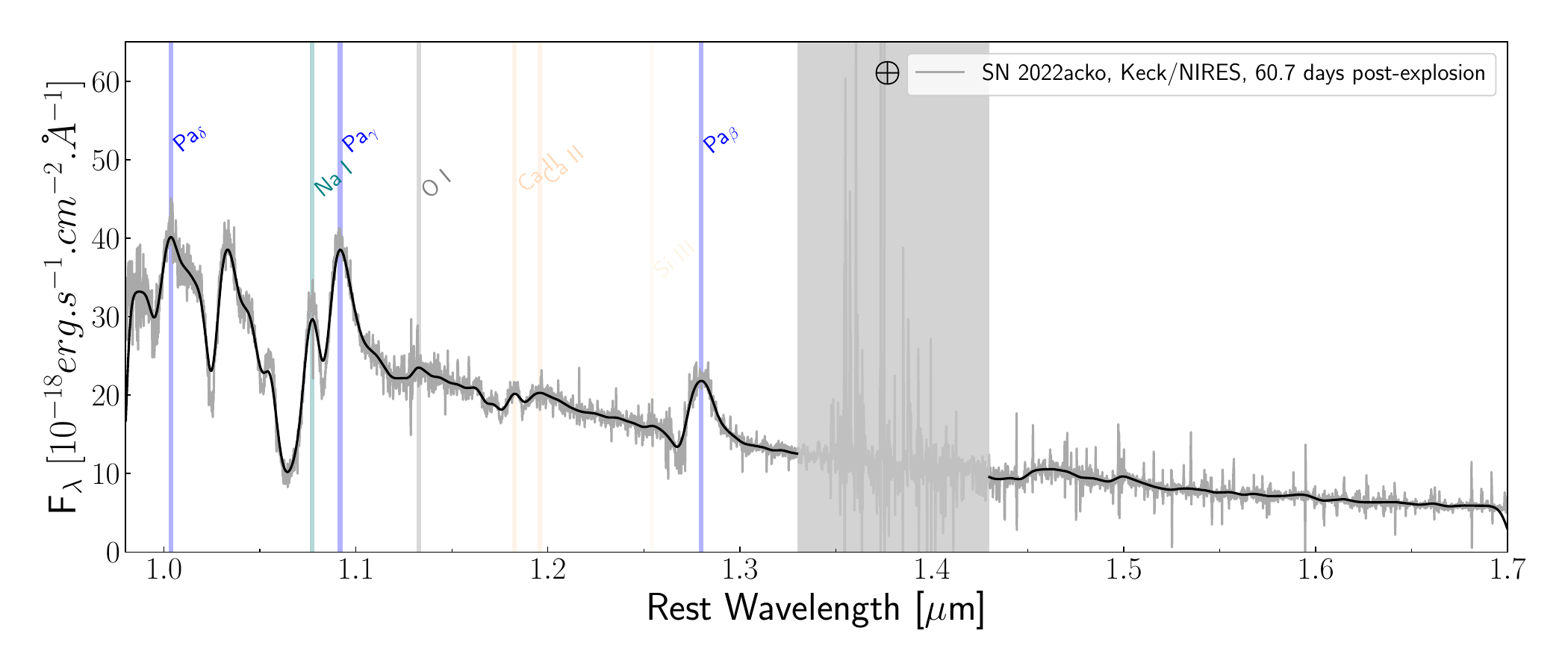}
    \caption{\textit{First:} NOT (+ ALFOSC) optical spectrum of SN~2022acko, with labeled line IDS.
    \textit{Second:} The Keck-II (+ NIRES) NIR spectrum  with labeled line IDs. 
}
    \label{fig:AllSpec1}
\end{figure*}

\begin{figure*}
    \centering
      \includegraphics[width=0.85\textwidth]{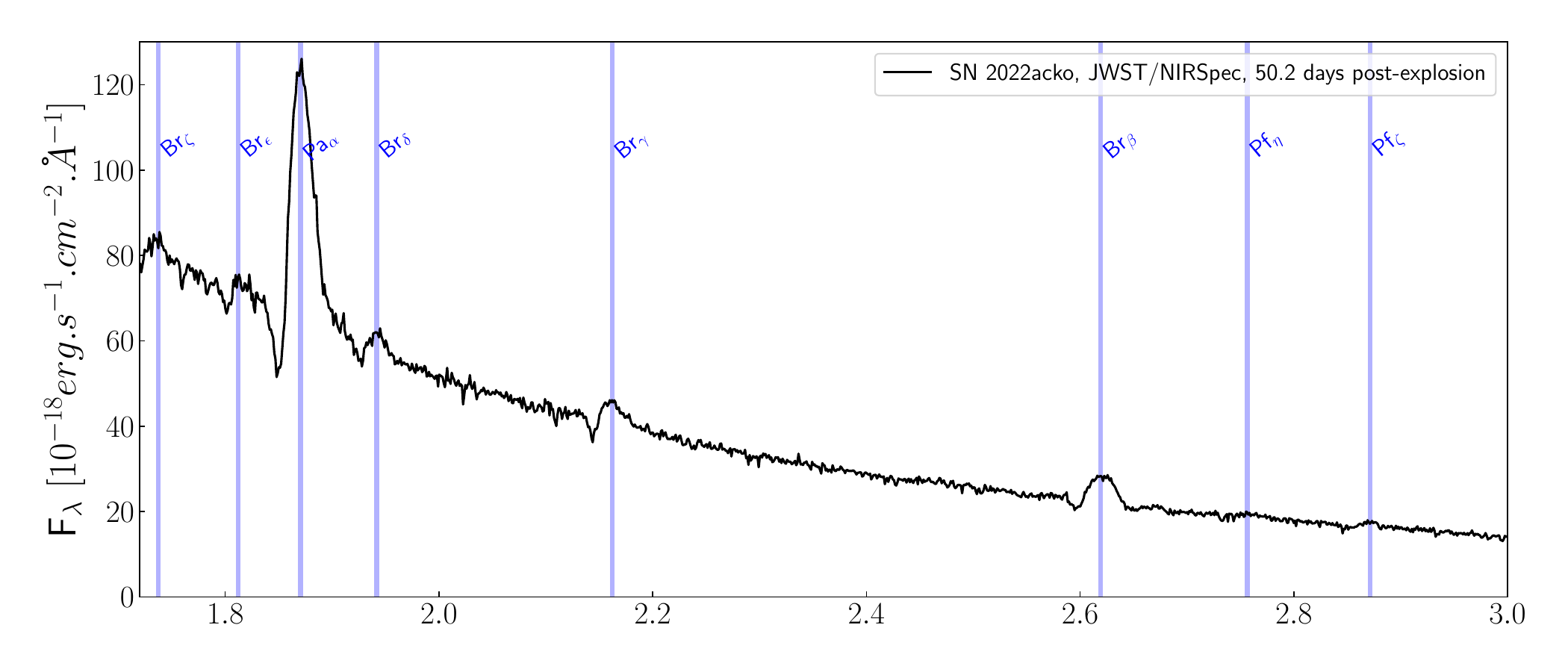}
       \includegraphics[width=0.85\textwidth]{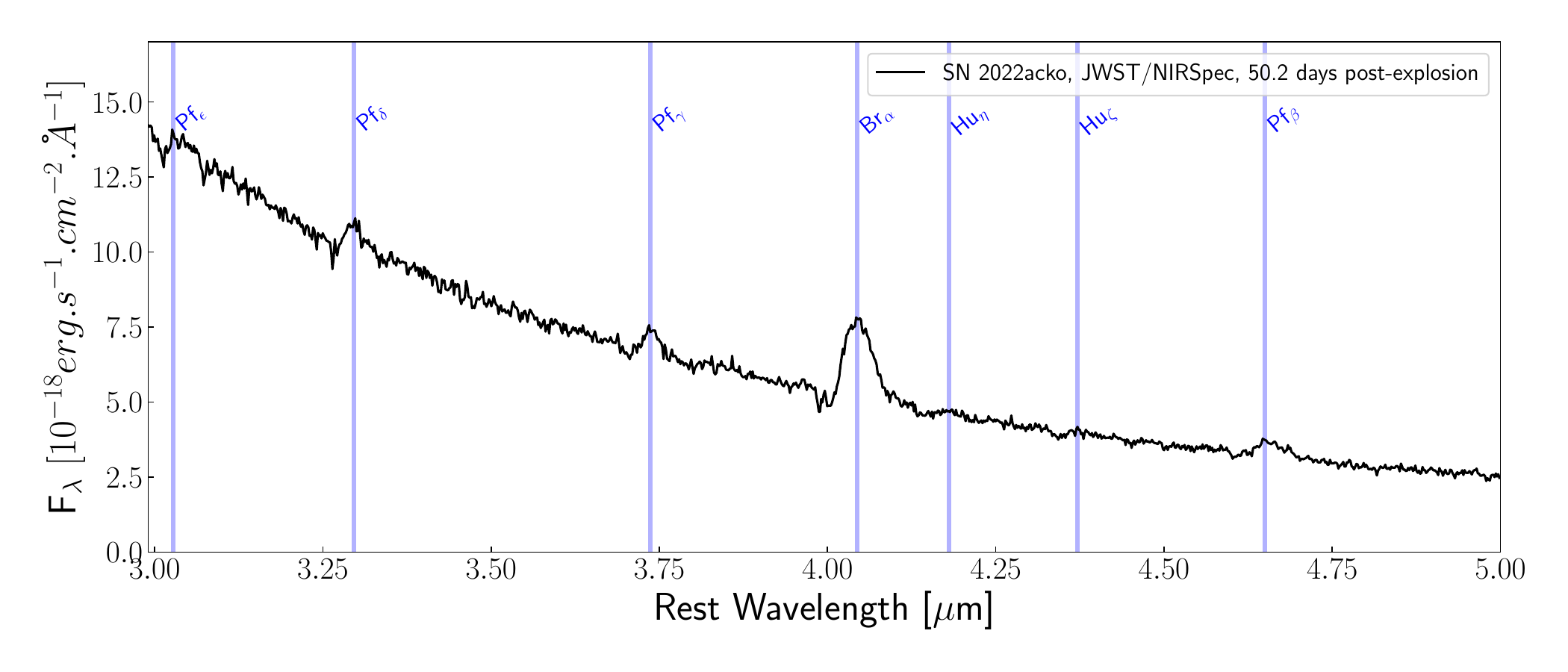}
    \caption{
    \textit{First:} The JWST-NIRSpec S400A1/F170P spectrum with labeled line IDs.
    \textit{Second:} The JWST-NIRSpec S400A1/F290P spectrum with labeled line IDs.}
    \label{fig:AllSpec2}
\end{figure*}

\begin{figure*}
    \centering
    \includegraphics[width=0.85\textwidth]{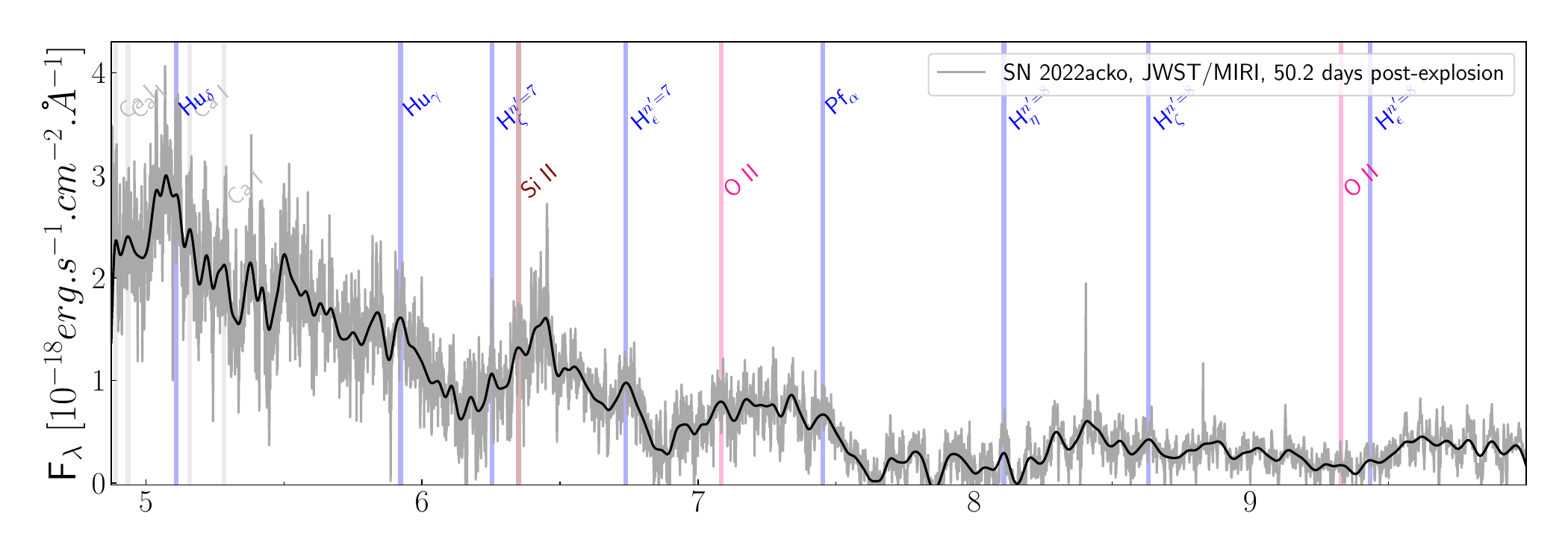}
    \vspace{0.5mm}
     \includegraphics[width=0.85\textwidth]{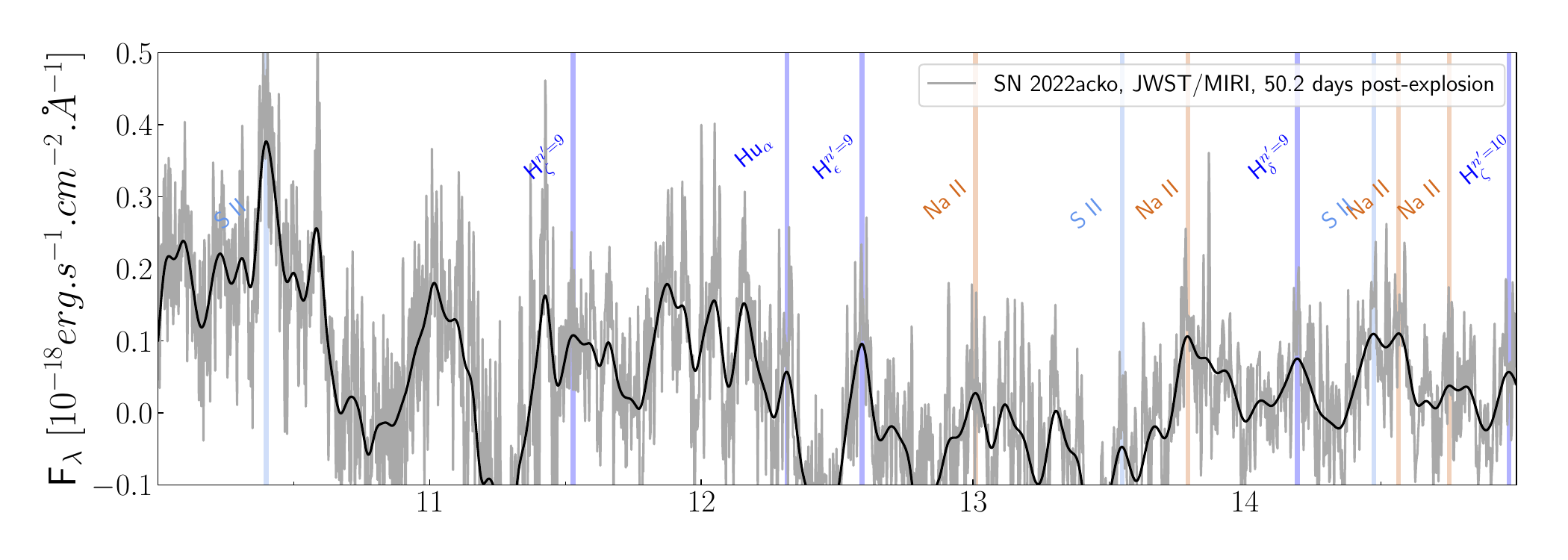}
     \vspace{0.5mm}
      \includegraphics[width=0.85\textwidth]{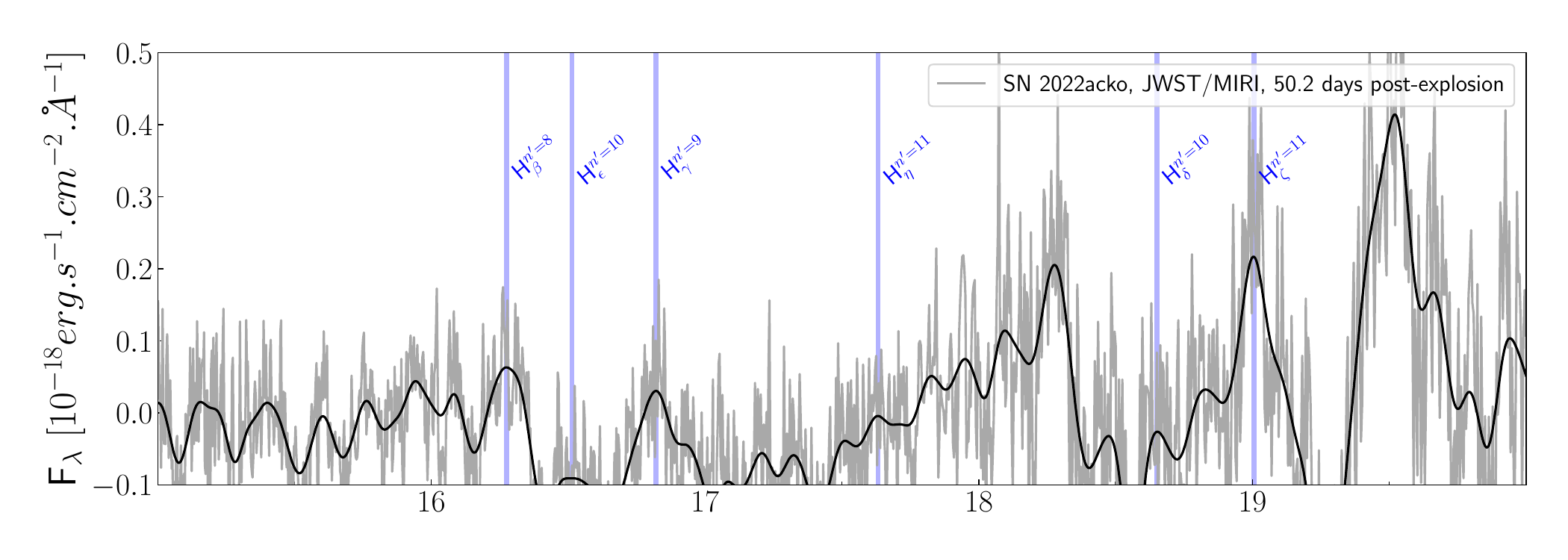}
      \vspace{0.5mm}
       \includegraphics[width=0.85\textwidth]{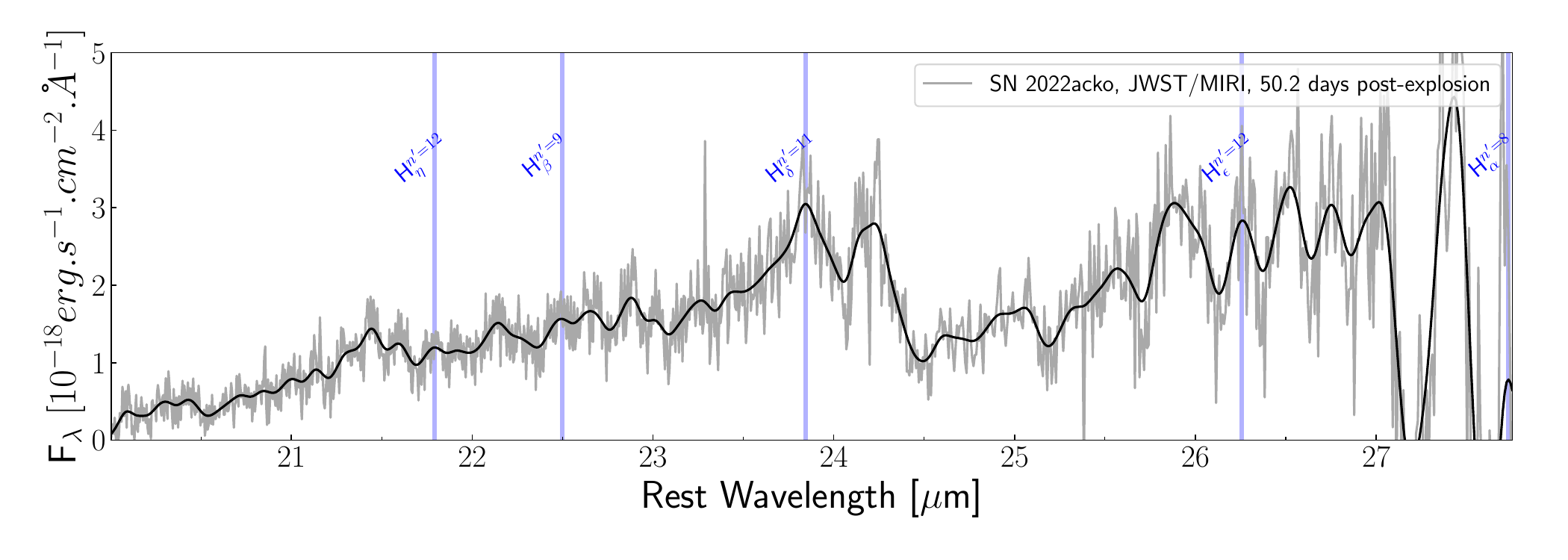}
       \vspace{0.5mm}
    \caption{The reduced stitch 4 channel MIRI-MRS spectrum of SN~2022acko with line identifications. Line identifications in Channel  4 are tentative due to the large background on the observations.}
    \label{fig:MIRIfullSpec}
\end{figure*}

\subsection{MIRI Medium Resolution Spectroscopy}
Medium resolution spectroscopy (MRS) data were obtained
with the MIRI instrument on \textit{JWST}. 
Observations were taken in the Short, Medium, and Long gratings to ensure continuous wavelength coverage (5$-$25~$\micron$) and stitched together.  
We note that the data longward of 15~$\micron$ is dominated by noise and instrumental thermal background; this is in part because of the scientific goals of these
observations were to cover the $SiO$ fundamental overtone at
$\sim$8.1~\um, the exposure time was not optimized for the 
longest wavelengths, and because of the known issues with the degrading count rate throughput of Ch3 and Ch4\footnote{https://www.stsci.edu/contents/news/jwst/2023/miri-mrs-reduced-count-rate-update}. 
The MRS data were reduced using a custom-built pipeline\footnote{\url{https://github.com/shahbandeh/MIRI_MRS}} 
designed to extract point source observations with highly varying and/or 
complex backgrounds from MRS data cubes. The pipeline 
creates a master background based on 48 different positions in the data cube. 
This master background is then subtracted from the whole data cube before the 
aperture photometry is performed along the data cube, using the \textsc{Extract1dStep} 
in stage 3 of the \textit{JWST} reduction pipeline. The reduction shown
here utilized version 1.12.1 of the JWST Calibration pipeline \citep{Bushouse2023_JWSTpipeline}
and Calibration Reference Data System files version 11.17.1. Figure \ref{fig:cubes} shows the stacked MRS data cubes in each channel before and after background subtraction, and the spectrum is shown in full in Figure~\ref{fig:MIRIfullSpec}. 
The raw data associated with this reduction can be found at
\dataset[DOI:10.17909/ea3g-5z06]{\doi{10.17909/ea3g-5z06}}.

\subsection{Ground-based spectroscopy}
To complete the SED to shorter wavelengths, we also present ground-based spectra of SN~2022acko at similar epochs. 
A summary of these optical and NIR observations is provided in Table~\ref{tab:22ackodetails}.

An optical spectrum was obtained with the NUTS2 collaboration using the 2.54-m Nordic Optical Telescope (NOT) equipped with the Alhambra Faint Object Spectrograph and Camera (ALFOSC). 
The  spectrum  was taken on MJD 59978.86 which corresponds to  +60.7~days post-explosion.  The data were reduced within the \texttt{PyRAF} environment using the NUTS2 pipeline written by E. Cappellero. 

A NIR spectrum was taken on MJD 59951.95 at +33.78~days post-explosion with the Near-Infrared Echellette Spectrometer (NIRES)  mounted to the 10-m Keck II telescope. 
\acko\ was observed at two positions along the slit (AB pairs) to perform background subtraction. 
An A0V star was observed immediately adjacent to the science observation, and used for both telluric and flux calibration. 
The NIRES data were reduced using the \texttt{PYTHON}-based \texttt{pypeit} package. 
Combining these four spectra produces the first-ever nearly-continuous spectral coverage between $\sim$~0.4--25~$\micron$, which demonstrates the power of combining ground-based and \textit{JWST} data.

Two additional spectra were taken using the ALFOSC on the NOT via the NUTS collaboration, and an additional ground-based NIR spectrum was taken with EMIR mounted on the Gran Telescopio Canarias (GTC). 
These data are included in the online supplementary material.

\section{Line Identifications and Data Comparsion} \label{sec:IDs}
The following section discusses the spectral lines contributing to the spectral formation.  
The spectra of SN~2022acko are taken towards the end of the plateau phase. 
The hydrogen (H) spectral series, including the Balmer, Paschen, Brackett, Pfund, and Humphreys dominate the spectra of SN~2022acko. 
We also see some features of heavier elements \citep[e.g.,][]{1992A&A...258..399M}. Table~\ref{tab:speclines} lists all of the spectral lines that have been identified in the full spectrum, and below, we discuss the line identifications within various regions. 

\subsection{Optical Spectrum}
The top panel of Fig. \ref{fig:AllSpec1} shows the optical spectrum of SN~2022acko. For assistance with line IDs we use the models from \citet{Baron03}. In the optical SN~2022acko is dominated by the  Balmer series with some Paschen series: H$_\alpha$ (0.6563~\micron), H$_\beta$ (0.4861~\micron), H$_\gamma$ (0.4340~\micron),
H$_\delta$ (0.4102~\micron), Pa$_\eta$ (0.901~\micron), Pa$_\zeta$  (0.923~\micron), and Pa$_\epsilon$ (0.955~\micron).
Other ions which contribute to the 
spectral line formation include: the \CaII\ H\&K lines
(0.3933~\micron, 0.3968~\micron) and \CaII\ NIR triplet (0.8498~\micron, 0.8542~\micron, 0.8662~\micron), \MgII\
0.9218~\micron, \NaI\ 0.5896~\micron, 0.5890~\micron,
\ScII\ 5527~\micron, 5698~\micron, 6280~\micron, \BaII\  6142~\micron,
and many lines of \FeII\ including 0.4359~\micron, 0.4642~\micron,
0.4862~\micron, 0.4924~\micron, 0.5018~\micron, and 0.5169~\micron.  As well as \ion{C}{1} lines
particularly the line at 0.94~\micron, just redward of Pa$_\zeta$.


\subsection{NIR ground-based spectrum}
We use the lines IDs from \citet{Davis_2019} and \citet{Shahbandeh_2022} to assist us with identifying the main lines which contribute to the ground-based NIR spectrum. 
With the exception of  Pa$_\alpha$, which is located in the NIRSpec data,
the H features in the ground-based NIR spectrum are dominated by the  Paschen series. These include Pa$_\beta$ (0.1282~\micron),  Pa$_\gamma$ (0.1094~\micron), and  Pa$_\delta$ (0.1005~\micron). Other lines identified in this spectrum are \SrII\ 1.0040~\micron, 1.0330~\micron, \OI\ 1.1290~\micron,  \MgII 1.0092, 1.0927~\micron, \CaII\ 1.1839~\micron, 1.1950~\micron, and \CI\ 1.0694~\micron, 1.1333~\micron, 1.175~\micron.

\subsection{NIR JWST/NIRSpec spectrum}
 Figure \ref{fig:AllSpec2} shows the NIRSpec spectrum obtained at $\sim$+50~d. The spectrum is dominated by P Cygni profiles of H lines mainly coming from the Brackett and Pfund series. The main lines in the spectra include   Br$_\epsilon$ (1.817~\micron), \HI\ 1.875~\micron,  Br$_\delta$ (1.944~\micron), Br$_\gamma$ (2.166~\micron), Br$_\beta$ (2.626~\micron), \HI\ 2.873~\micron, 3.297~\micron, 3.741~\micron, Br$_\alpha$ (4.052~\micron), \HI\ 4.377~\micron, Pf$_\beta$ (4.653~\micron), \HI\ 5.129~\micron.

\subsection{MIR JWST/MIRI spectrum}
Figure~\ref{fig:MIRIfullSpec} shows the lines identified in the MIRI spectrum.
The MIRI spectrum is comprised of observations across four channels, which range from 4.90-7.65~\micron\, for Channel 1, 7.51-11.70~\micron\, for Channel 2, 11.55-17.98~\micron\, for  Channel 3, and 17.70-27.90~\micron\, for Channel 4. 
Visual inspection of the non-background subtracted data cubes shows there is clear SN flux in the first two channels; the singal-to-noise (S/N) of the SN varies between 20 and 2
from 4.90-11.70~\micron, after background subtraction, this S/N
increases to 35 to 6 (see Figure~\ref{fig:cubes}). 
This gives us confidence that our line identifications shortwards of 11.70~\micron\ are accurate. 
Before background subtraction, in Channel 3, the S/N varies from 1-2 depending on the exact wavelength range. After background subtraction, the S/N is $\sim$3.  
Due to the high background and low throughput of MIRI-MRS in channel 4, visual inspection of the pre-background subtracted data shows no sign of the SN. However, after background subtraction, (i.e. when the data is fully reduced using our pipeline) the supernova is  visible in Channel 4 at a S/N of 1-3.
For example, we clearly see the SN at wavelength 22.167~$\micron$, 25.1854~$\micron$, 26.9623~$\micron$. Therefore, we still choose to identify possible spectral lines in this region but as we could not manually inspect every wavelength slice for possible SN signal, we want to remind the reader that the identifications in this Channel are tentative. 

Generally, the MIR spectrum is dominated by H lines (transitions from levels 6, 7, 8, and 9), however, there are
several features that are from other ions. Some of the most notable
spectral features include \CII\ 5.539, 5.836, 6.775, 6.900~$\micron$,
\NII\ 8.439, \NIII\ 9.175, 9.191~$\micron$, \OI\ 7.720,
10.384~$\micron$ and \OII\ 9.332, 9.439~$\micron$. The observation of
s-process elements and lines of C, N, and O in the
hydrogen elements validate our understanding of CNO burning during
stellar evolution.

\begin{deluxetable}{cccccc}
\tablecaption{SN~2022acko spectral lines.  \label{tab:speclines}} 
\tablehead{
\colhead{Ion} & 
\colhead{Wavelength}& 
\colhead{Ion} & 
\colhead{Wavelength}&
\colhead{Ion} & 
\colhead{Wavelength}\\
 & 
 \colhead{[$\mu$m]} &  
 & 
 \colhead{[$\mu$m]}& 
 & 
 \colhead{[$\mu$m]}}
\startdata
\multicolumn{6}{c}{Optical spectrum}  \\
\hline
\HI   & 0.4861  & \ScIII & 0.6238 & \ScIII & 0.7449  \\
\HI   & 0.3889  &  \HI	 & 0.6563 & \HI	& 0.9229\\
\CaII & 0.3969  & \FeII & 0.4924 & \CaII & 0.8498 \\
\HI   & 0.3970  & \FeII & 0.5169 & \CaII & 0.8498 \\
\CaII & 0.3934  & \ScII & 0.5527 & \CaII & 0.8542 \\
\HI   & 0.4102  & \ScII & 0.5698 & \CaII & 0.8662 \\
\HI   & 0.4341  & \OI & 0.7774  & \MgII & 0.9218 \\
\FeII & 0.4358  & \NaI & 0.5894  & \CI   & 0.9408 \\
\FeII & 0.4642  & \BaII & 0.6142 & \HI	  & 0.9014\\
\hline
\multicolumn{6}{c}{Ground-based NIR spectrum }  \\
\hline
\HI   & 1.005  & \MgII & 1.0092  &  \OI  & 1.1290 \\
\HI   & 1.0938 & \MgII & 1.0927  & \SrII& 1.0330 \\
\HI   & 1.2818 & \NaI   & 1.0749 & \SrII& 1.0920  \\
\CaII & 1.1839 & \CI   & 1.1333      &  \\
\CaII & 1.1950 & \CI   & 1.1756      &  \\
\hline
\multicolumn{6}{c}{NIRSpec spectrum }  \\
\hline
\HI  &  1.817  &  \HI  &  2.626 & \HI  &  4.052 \\
\HI  &  1.875  &  \HI  &  2.873 & \HI  &  4.377 \\
\HI  &  1.944  &  \HI  &  3.297 & \HI  &  4.654 \\
\HI  &  2.166  &  \HI  &  3.741 & \HI  &  5.129\\
\hline
\multicolumn{6}{c} {MIRI/MRS spectrum}\\
\hline
\HI  & 5.129   & \OI  & 10.384  & \FeI & 16.596 \\
\CII & 5.539   & \OI  & 10.399  & \HI  & 16.881 \\
\HI  & 5.711   & \HI  & 11.309  & \HI  & 18.615 \\
\CII & 5.836   & \OI  & 11.853  & \NI  & 18.811 \\
\HI  & 6.291   & \OI  & 12.058  & \HI  & 18.978 \\
\HI  & 6.772   & \HI  & 12.372  & \HI  & 19.062 \\
\CII & 6.775   & \HI  & 12.390  & \NI  & 19.616 \\
\CII & 6.900   & \OI  & 12.995  & \NI  & 19.641 \\
\HI  & 7.460   & \CI  & 13.297  & \NI  & 20.353 \\
\OI  & 7.720   & \CI  & 13.370  & \NI  & 21.022 \\
\CI  & 8.374   & \CI  & 13.872  & \CI  & 22.278 \\
\NII & 8.439   & \HI  & 13.942  & \HI  & 23.868 \\
\HI  & 8.760   & \HI  & 14.183  & \NI  & 25.058 \\
\NIII& 9.175   & \HI  & 14.962  & \HI  & 26.168 \\
\NIII& 9.191   & \FeI & 15.561  & \HI  & 26.682 \\
\OII & 9.332   & \CI  & 15.803  & \CI  & 27.425 \\ 
\HI  & 9.392   & \FeI & 16.107  & \HI  & 27.803 \\ 
\OII & 9.439   & \HI  & 16.209   \\
\hline
\enddata
\end{deluxetable}

\subsection{Spectral Comparison}

\begin{figure*}
    \centering
    \includegraphics[width=0.99\textwidth]{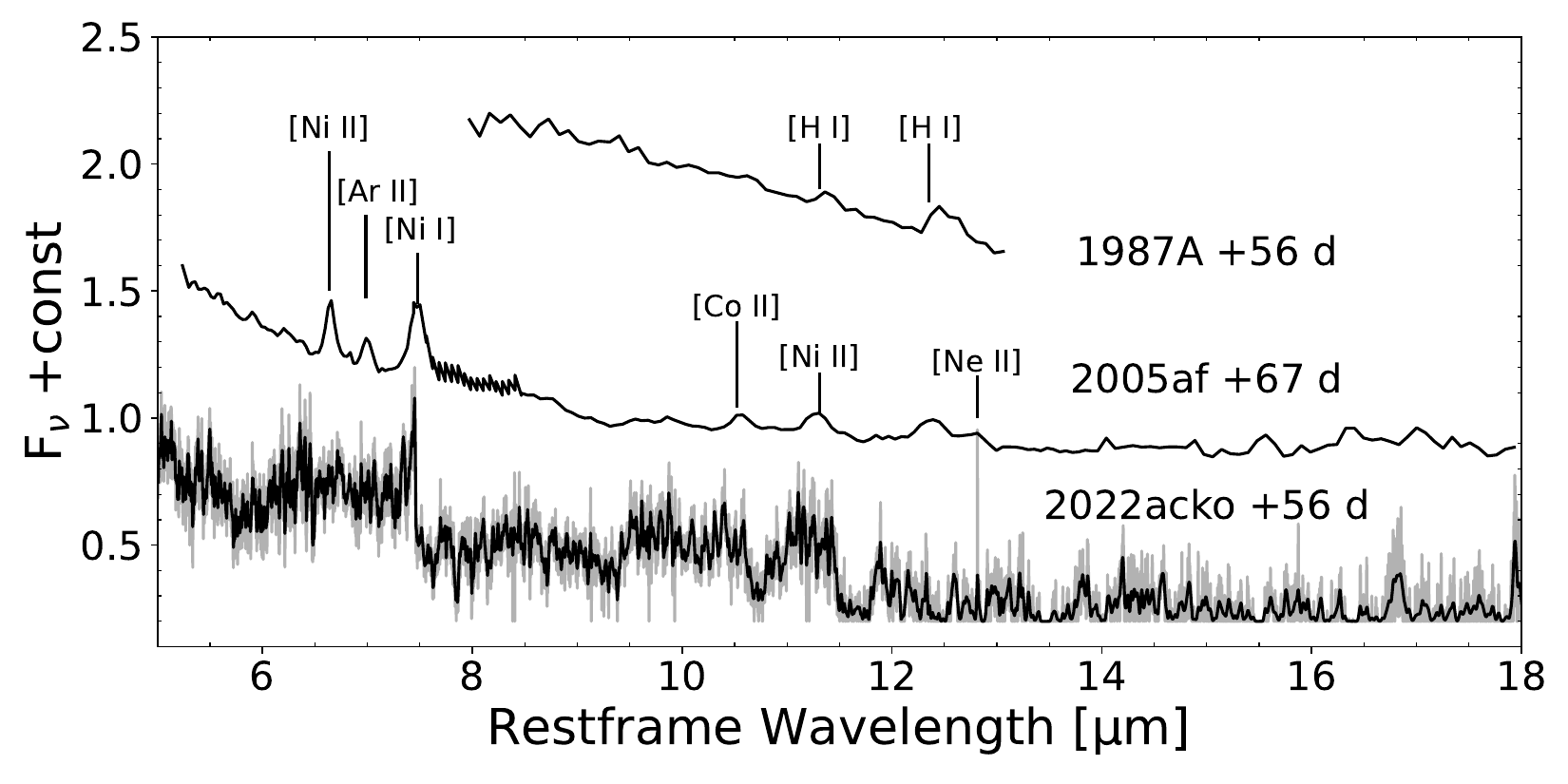}
    \caption{Comparison between MIR spectra of SN 1987A, 2005af and 2022acko.  the forbidden lines seen in the day 67 spectrum of SN 2005af are absent in earlier spectra of SN 1987A and 2022acko. This is because at these phases SN~2005af was at the end of the plateau. the data from SN~1987A are digitalized from \cite{Aitken88,Bouchet89}.}
    \label{fig:mir_copmp}
\end{figure*}

In this section, we examine the differences between SN~2022acko and other CCSN which have MIR spectra. Figure~\ref{fig:mir_copmp} shows a comparison of the Spitzer Space Telescope data of SN\,2005af  obtained at day~+67 \citep{Kotak_etal_2006_05af} with that of SN~1987A at day +56 \citep{Aitken88}  and the MIRI/MRS data of SN~2022acko obtained at day~+56~d. 
The three SNe are generally similar  
but have a few notable differences. 
At this epoch, SN~2022acko is still on the 
plateau phase and is dominated by H features, whereas  SN~2005af was faster evolving and passed the plateau phase by day +67 \citep{Kotak_etal_2005_04dj}. 
In fact,  the forbidden lines 
identified by \citet{Kotak_etal_2006_05af} are absent in both
 SN~1987A and SN~2022acko.
Although the SN~2022acko MIRI spectrum is $\sim 10$~days earlier than SN~2005af, it appears to be evolving more slowly, thus we would expect the ZAMS mass of SN~2022acko to be larger than that of SN~2005af.
SN~2004dj was also observed by the \textit{SST} at epochs of 89--129 days \citep{Kotak_etal_2005_04dj}.
In this spectra  [\NiII] lines were seen and the presence of $CO$ was inferred as early as 106~days. Future observations of SN~2022acko will allow us to compare the epoch of $CO$ formation, and evolution of the onset of forbidden lines, which will enable us to determine if SN~2022acko had a larger ejecta mass.

\section{Hydrogen Velocities}\label{sec:Hvel}
\begin{figure*}
    \centering
    \includegraphics[width=0.33\textwidth]
    {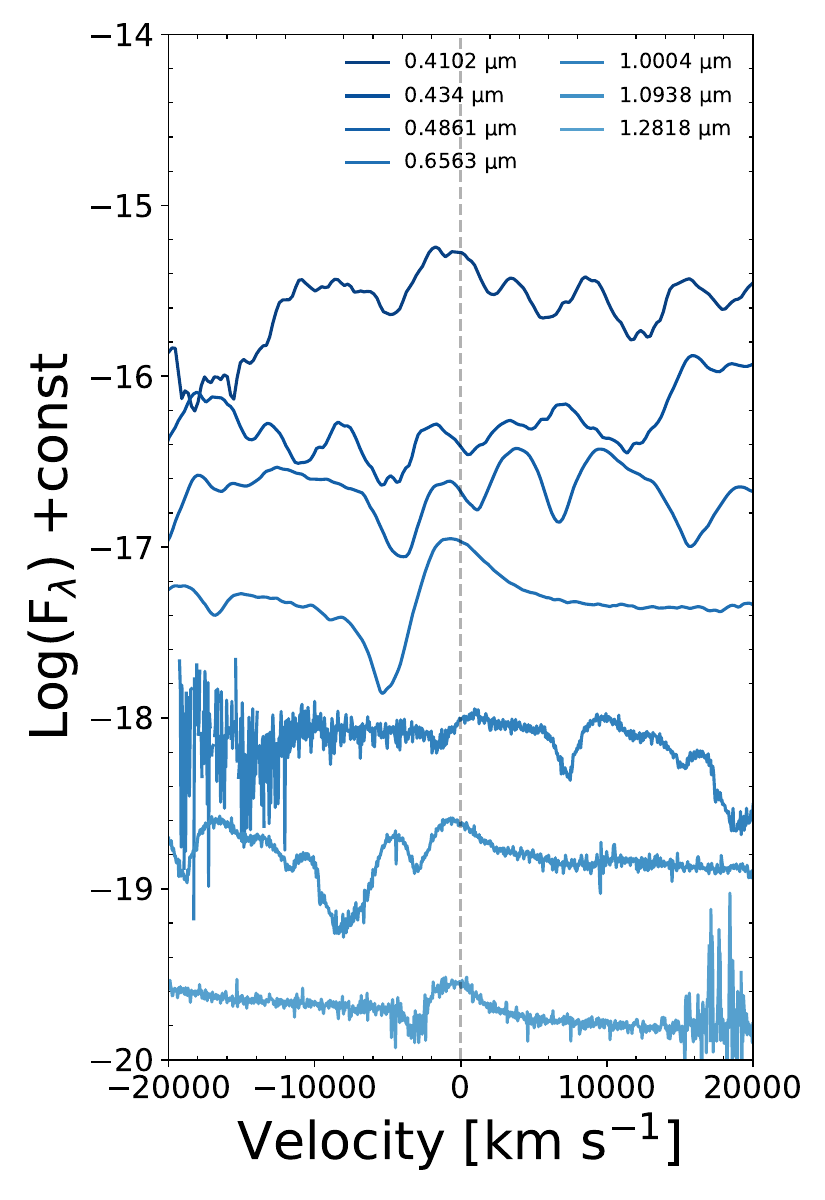}
    \includegraphics[width=0.33\textwidth]
    {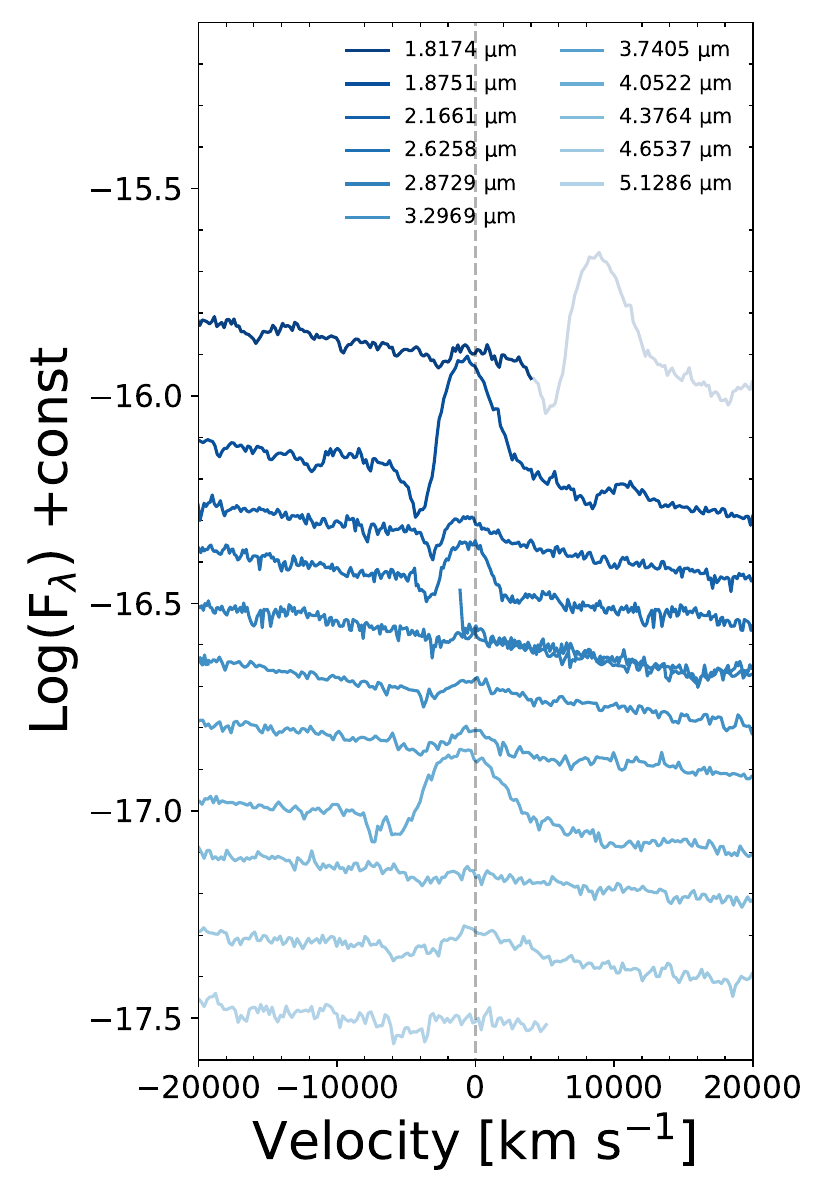}
    \includegraphics[width=0.33\textwidth]
    {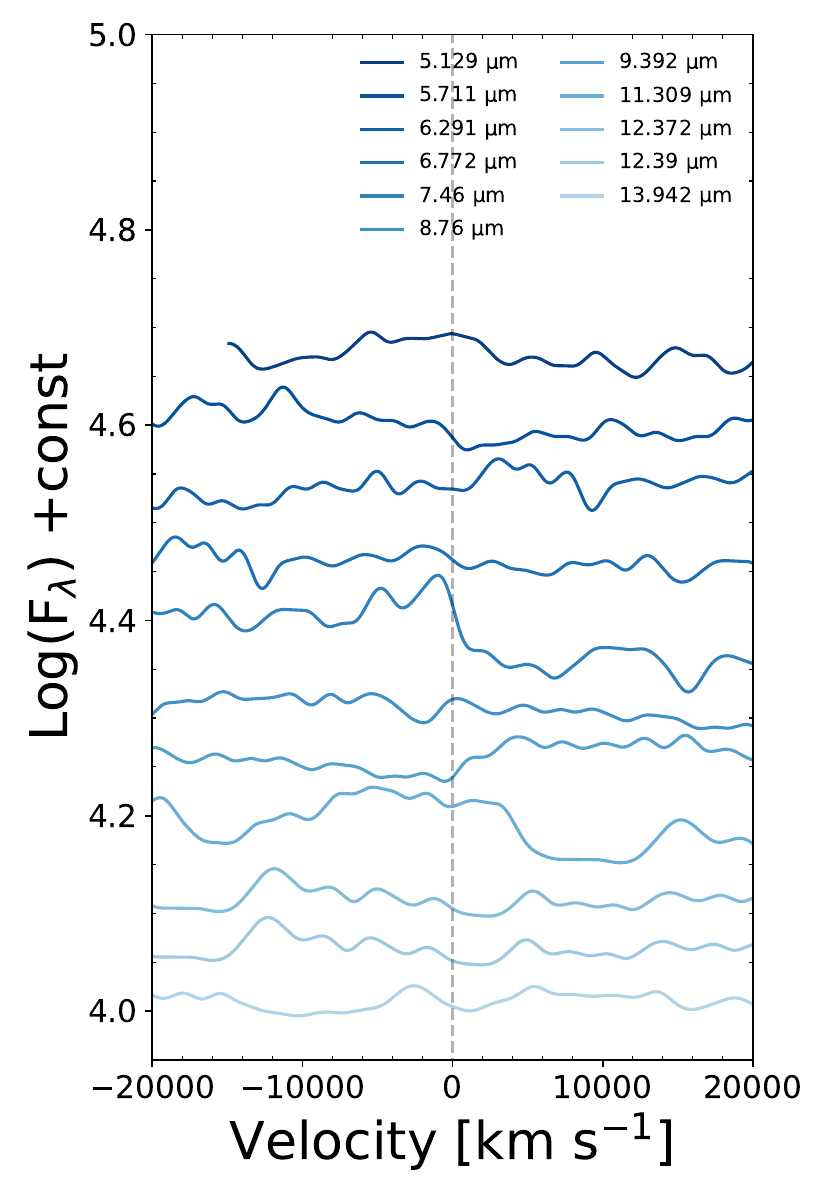}
    \caption{The H line velocity profiles from the optical to the MIR wavelengths. We identify 29 potential H features in the ground-based optical and NIR spectra (\textit{left}), the JWST NIR spectra (\textit{middle}), and the JWST mid-IR spectra (\textit{Right}).    }
    \label{fig:velprofiles}
\end{figure*}

\begin{figure*}
    \centering
    \includegraphics[width=0.98\textwidth]{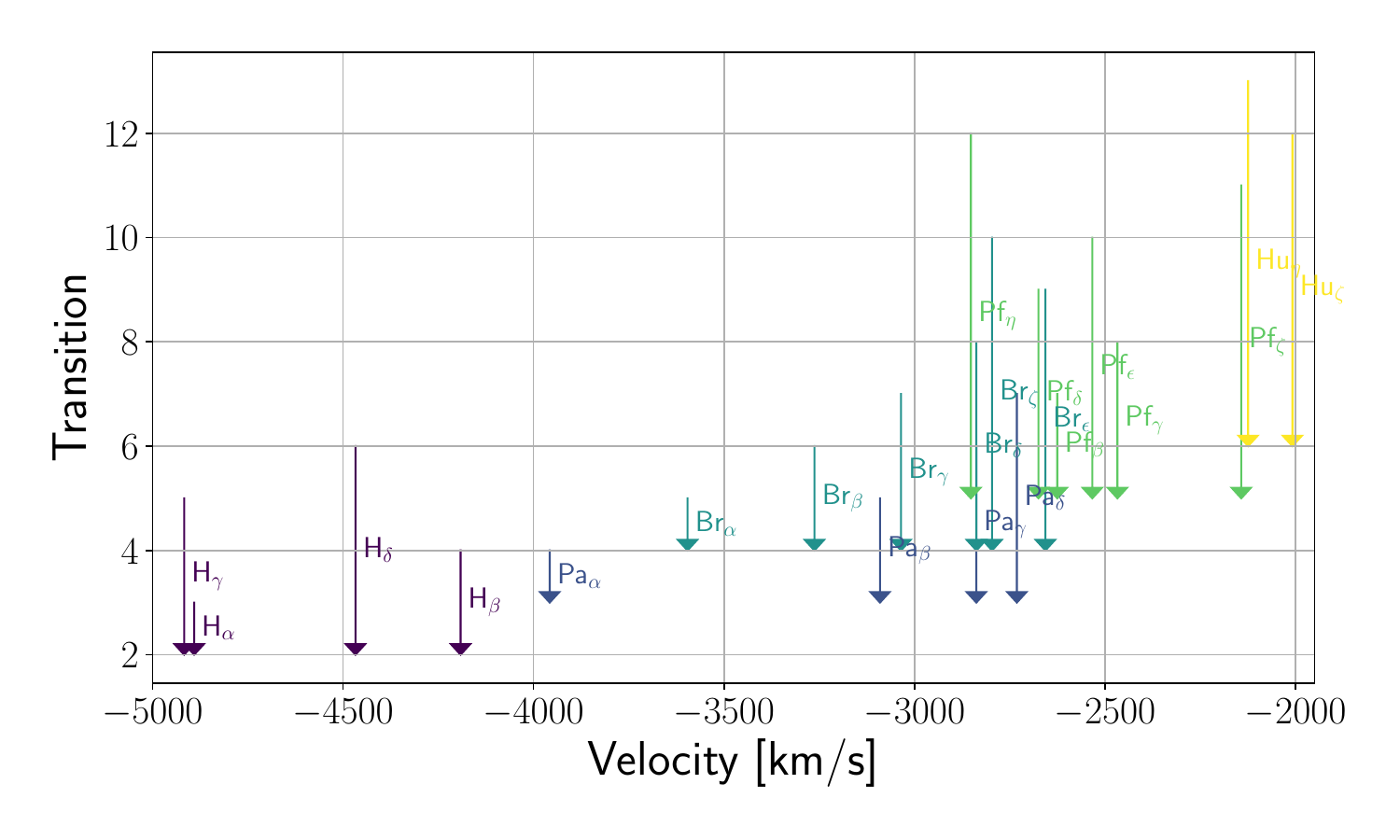}
    \caption{The minimum velocity of each hydrogen feature against the transition at which it occurs.} 
    \label{fig:abs_vels}
\end{figure*}

To understand where in the ejecta the line formation of H is occurring we examine the profiles of all of the H lines in velocity space. 
Figure \ref{fig:velprofiles} shows the velocity profiles of Hydrogen from 0.4-25~$\micron$. 
Overall, we see evidence for $\sim$30 different spectral lines. Where four are from the ground-based optical spectrum, three are from the ground-based NIR, 11 are from the JWST NIRSpec data, and 11 are from the JWST MIRI spectrum. 
To determine the velocity of each absorption, each H-profile was fit with a Gaussian function using the method described in \citep{Shahbandeh_2022}. 
The H absorptions vary in velocity from  2100 to 4900~\kms; see Table \ref{tab:speclines}.
The velocities of the H lines are strongly dependent on the transition, whereas the lower atomic transitions have increasingly higher velocities, see Fig. \ref{fig:abs_vels}. We also note that for all the H P-Cygni features, the emissions of the profiles are blue-shifted by 1000-2000~\kms; this is to be expected because the extended atmosphere will block parts of the red-shifted emission.

\begin{deluxetable}{lccc}\label{tab:abs_vels}
  \tablecaption{H I absorption velocities.  Spectral lines affected by significant blending are indicated with an *. \label{tab:mir_lines} } 
  \tablehead{\colhead{Species} & \colhead{Rest Wavelength [$\mu$m]} & \colhead{Velocity [\kms]} & \colhead{Error [\kms]}}
  \startdata
H$_{\gamma}$ & 0.434 & $-$4900 & 900 \\
H$_{\beta}$ & 0.486 & $-$4200 & 900 \\
H$_{\alpha}$ & 0.656 & $-$4900 & 900 \\
Pa$_{\eta}$* & 0.901 & 400 & 900 \\
Pa$_{\zeta}$* & 0.923 & $-$2300 & 900 \\
Pa$_{\epsilon}*$ & 0.955 & $-$2400 & 900 \\
Pa$_{\delta}$ & 1.005 & $-$2700 & 200 \\
Pa$_{\gamma}$ & 1.094 & $-$2800 & 100 \\
Pa$_{\beta}$ & 1.282 & $-$3100 & 100 \\
Pa$_{\alpha}$ & 1.876 & $-$4000 & 300 \\
Br$_{\zeta}$ & 1.737 & $-$2800 & 400 \\
Br$_{\epsilon}$ & 1.818 & $-$2700 & 400 \\
Br$_{\delta}$ & 1.945 & $-$2800 & 400 \\
Br$_{\gamma}$ & 2.166 & $-$3000 & 300 \\
Br$_{\beta}$ & 2.626 & $-$3300 & 300 \\
Pf$_{\eta}$ & 2.758 & $-$2900 & 300 \\
Pf$_{\zeta}$ & 2.873 & $-$2100 & 400 \\
Pf$_{\epsilon}$ & 3.039 & $-$2500 & 300 \\
Pf$_{\delta}$ & 3.297 & $-$2700 & 400 \\
Pf$_{\gamma}$ & 3.741 & $-$2500 & 400 \\
Br$_{\alpha}$ & 4.052 & $-$3600 & 400 \\
Hu$_{\eta}$ & 4.171 & $-$2100 & 400 \\
Hu$_{\zeta}$ & 4.376 & $-$2000 & 400 \\
Pf$_{\beta}$ & 4.654 & $-$2600 & 400 \\
\enddata
\end{deluxetable}

\section{Full Spectral Energy Distribution}\label{sec:SED}
\begin{figure*}
    \centering
    \includegraphics[width=\textwidth]{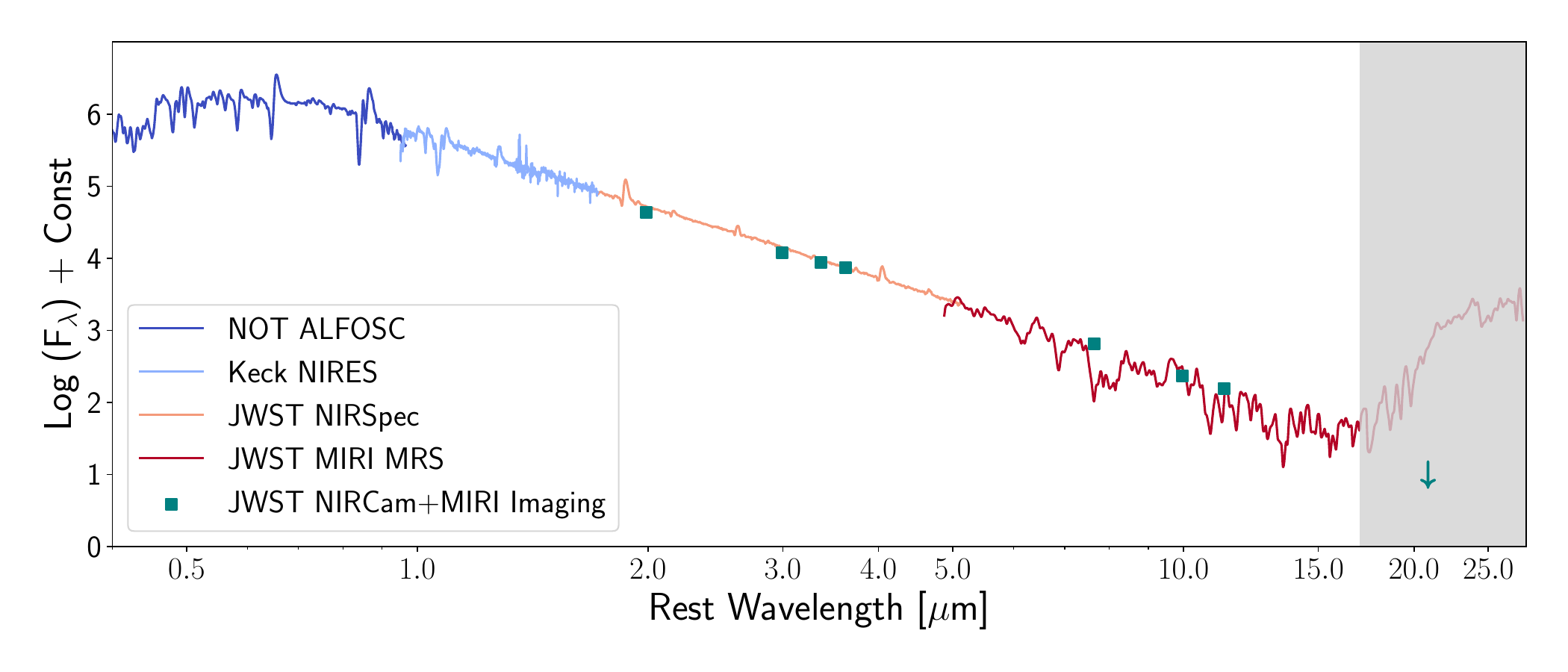}
    \caption{The spectral energy distribution of SN~2022acko extending between 0.4--25~$\mu$m. A constant has shifted the Keck (+ NIRES) spectrum to match the optical NOT (+ ALFOSC)  and the \textit{JWST} spectra.  The NIRSpec and MIRI fluxing is consistent with the multi-band \textit{JWST} photometry, except in Channel 4, where the background dominates the spectral flux. The \textit{F2100W}-band photometry sets an important base level for future \textit{JWST} observations. SN~2022acko fit with a blackbody function characterized by $T_{eff} =4000~K\pm500$ is shown in golden.}
    \label{fig:fullSED}
\end{figure*}

By combining all of the spectra, we can produce the first 0.4 to 25~\micron\ SED of a SN~IIP, see Fig. \ref{fig:fullSED}.  The average time of the SED is at $\sim52$~d past explosion. For orientation, we characterize the overall energy distribution by the temperature of the black body. This is possible because Thomson and free-free emission dominate the NIR and MIR continuum. When fit with a blackbody function, the SED produces a temperature of $\sim$4000~K.

We compared the \textit{JWST} photometry with  \acko's SED and found the NIRspec and MIRI spectra to be consistent with the photometry, except for wavelengths longer than 18~$\micron$, where the MIRI spectrum increases in flux. The background in Channel 4 causes this increase in the spectral flux and is not intrinsic to the SNe or environment. Hence, we see no evidence of pre-existing dust or molecules in the SN~2022acko, and this statement will be quantified below.

\section{Limits on CO formation and pre-existing molecules and dust}
\label{sec:COformation}

  To obtain upper limits on the $CO$ formation, we use our framework MOLFIT which provides an interactive layer based on the module for calculating the vibrational and rotational bands of molecules and for spherical radiation transport in HYDRA  (see \citep{1989HiA.....8..207S,Hoeflich_1988,Gerard_etal_2000,Hoeflich_2002,Rho21}, and references therein). 
  To reproduce the observations the following free parameters are used: a) temperature, b) density represented by a power law $\propto \rho ^n$, c)
  abundances, d)  small-scale random motion such as turbulent motion  $v_\text{turb}$, e) average expansion
  velocity $v$ of a shell of thickness $\delta v$, and f) the $CO^+/CO$ ratio to mimic the possible exposure to hard $\gamma $-rays. 
  Rather than solving the time-dependent rate equations for molecule formation and radiation transport in HYDRA, local-thermodynamical equilibrium is used for the molecule to atomic abundances. 
  Note that the $CO$ concentration in LTE is an upper limit for a rapidly 
  expanding atmospheres because the formation times
  scale for $CO$ are comparable to geometrical dilution by expansion,
  with an additional role of hard radiation as discussed \citep[see,
  for example][]{Rho21}.
  It is well known that for the formation and destruction, time-dependent rate equations for
  $C+O \rightarrow CO$,\, $C^++O \rightarrow CO^+$ are needed. The
  exposure of the ejecta to $\gamma $-rays by $^{56}$Ni mixing
  may increase the $CO$ formation unless the resulting destruction
  outweighs the formation rate via the $C+O^+\rightarrow CO^+$
  channel, and that the thermodynamic equilibrium abundance is larger
  than the $CO$ abundances obtained by the time-dependent rate equations \citep{Spyromilio_1988,Sharp_1990,1990ApJ...358..262L,1993MNRAS.261..535M,Rho_2019_ATEL}. 
  Time-dependent rate equation effects have been included in models
  for SN~1987A and stripped-envelope SNe  \citep{1989HiA.....8..207S,
    Sharp_1990,Gerardy_etal_2000}. Including  these effects would
  result in a
  smaller upper limit of the $CO$-mass given below. 
  For the simulations, 50,000 to 300,000 vibrational transitions have been taken into account for each $CO$, $CO^+$ and $SiO$. 

\subsection{CO formation in the SN-ejecta}
The spectrum of SN~2022acko does not show evidence for the
formation of molecules. However, our data  provides important
upper limits on the amount of pre-existing molecules and dust,
which is required to set a baseline for observations at later times.
In addition, it probes the  mixing of C/O rich
material into the H-rich envelope. 
It is well-established for
CCSNe with a massive H-rich envelope, that molecule formation starts when the C/O core gets exposed
or when C/O plumes  are mixed into the H-rich envelope through  rayleigh taylor instabilities. This also only occurs when the  temperature drops below  $T \approx 5000$ K
\citep{Liu_Dalgarno_1995,Gerardy_etal_2000}. 
For example, in SN~1987A, there was 
 observational evidence for mixing of C/O up to $v \approx 3000$~\kms 
and H-rich material down to $v \approx 800$~\kms
\citep{1987ESOC...26..147L,1987MNRAS.229P..15C,Hoeflich_1988}. Mixing
has been discussed for other SNe II \citep{Gerardy_etal_2000,
  Kotak_etal_2005_04dj,Meikle_etal_2007,2009ApJ...704..306K,Rho_etal_2018,
  Davis_2019}. This mixing of C/O rich material was
understood as due to Rayleigh-Taylor instabilities developing in the ejecta between different chemical layers \citep{1989A&A...220..167M,Fryxell_etal_1991,1998ApJ...495..413N}.

For SN~2022acko, to obtain upper limits for $CO$ during the plateau phase, 
we assumed a density structure and temperatures, $T$, typical for a SN~IIP that
has an effective temperature $T \approx 5500$~K and a rapid drop in
$T$ well below 5000~K to about 2500--4000~K as the photosphere recedes
\citep{1993MNRAS.265..471T,2003MNRAS.345..111C,2013MNRAS.433.1745D}.
For a specific example, see Figs. 12--13 in \citet{2003MNRAS.345..111C}.  
 For the density gradient, we use the ratio of hydrogen absorption
 line shifts within each series of H-line features which is limited by the optical depth of 1 in the line where each member of a series has the same lower level or the optical depth of 1 in Thomson scattering, as previously employed for
 e.g. SN~1987A \citep{Hoeflich_1988}, and SN~1998S
 \citep{Gerardy_etal_2002}. 
 Comparing the velocities obtained for transitions with the same lower
 level from Table \ref{tab:abs_vels} and
 Fig.~\ref{fig:abs_vels} and assuming this corresponds to the point
 where $\tau =1 $ for this transition,  we obtain  $2.1 < n < 2.6$
 assuming the density follows a powerlaw, $\rho \propto r^{-n}$  \citep{Hoeflich_1990_Habil,Duschinger_etal_1995}.

 During the plateau, the location of the photosphere is determined by
 the recombination of H, because the opacities drop by 2 to 4 orders
 of magnitude   \citep[see, for example][their Figure 13]{2003MNRAS.345..111C}.

\begin{figure*}
    \includegraphics[width=0.99\textwidth]{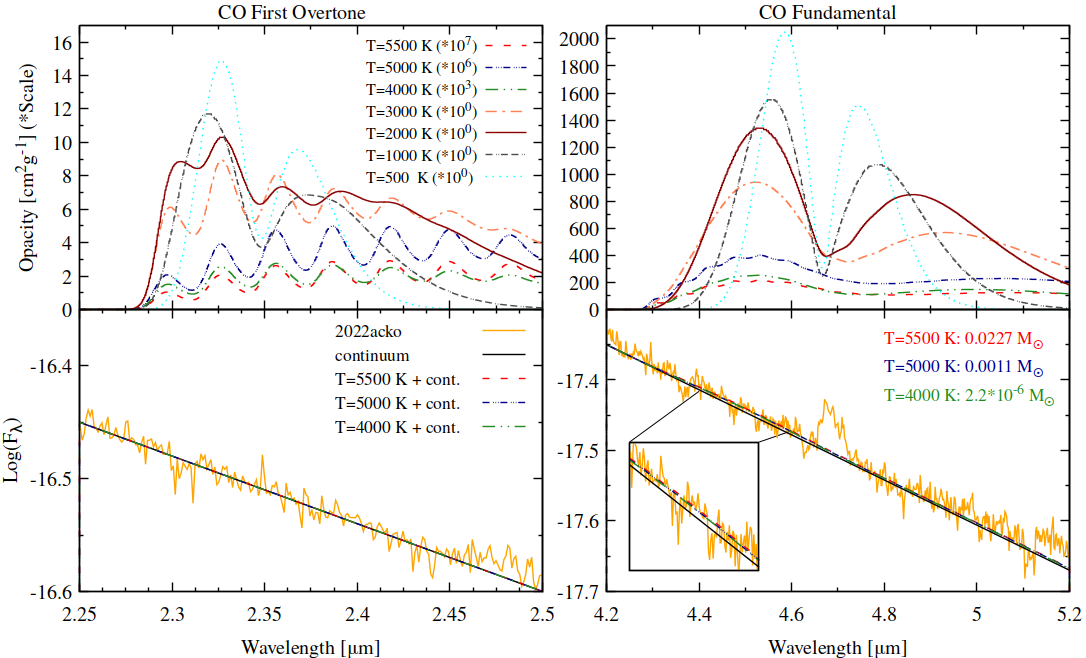}
    \caption{Limits of $CO$ in SN~2022acko. $CO$ opacities  cm$^2$~g$^{-1}$~\AA\ with scale factors of the  $1^{st}$ 
    (upper left) and fundamental band (upper right) of $CO$, and the comparison of the continuum + $CO$ emission and observation with the
    total $CO$ masses based on the fundamental band.  The specific $CO$-emissivity is given by the product of 
    opacity and black body.
    For the upper limit of the $CO$ emission, we assumed the S/N as limit for the relative flux contribution.  The underlying continuum the $CO$ feature is approximated akin a Blackbody. The upwards slope beyond 4.8 \mic hints that free-free emission is present. 
    Note that the mass limits based on the $1^{st}$ $CO$ would be larger by a factor of $\approx 100$, clearly demonstrating the need for \textit{JWST} observations.}
    \label{fig:CO_fit}
\end{figure*}

  Fig.~\ref{fig:CO_fit} shows the scaled opacity (top) for both the
  fundamental (right) and first overtone (left) of $CO$, and their
  flux for given temperatures comparison with observation
  (bottom). The opacity peaks at about 2500--3000~K and decreases by
  about six orders of magnitude towards the recombination temperature
  of H. 
  For lower temperatures, the flux in the fundamental band
  scales $\propto T$ as expected, whereas the first overtone flux
  drops explaining why this band vanishes at about day 200 in SNe~IIP
  \citep{Gerardy_etal_2000}. Because the opacity and  the specific
  emissivity relative to the continuum flux in the first $CO$
  overtone is smaller than the fundamental band by a factor on the
  order of $\approx10^2$, the fundamental band provides the upper
  limits, showing the importance of \textit{JWST} data.  
  
  At the inner edge of the photosphere ($T\approx 5000-5500$~K), when compared to an outer region ($T\sim 4000$~K), the $CO$ opacity is reduced by factors of three to four orders of magnitude, thus providing a sharp inner boundary for the $CO$ emission. 
  For $T=4000$~K, we obtain $M(CO)\approx  2 \times 10^{-6} M_\odot$ 
  (Fig. \ref{fig:CO_fit}, lower right), under the reasonable assumption that the mixing of C/O rich material does not stop exactly at the photosphere. We have
 divided our upper limit by the signal to noise ratio (S/N $\approx$44) To obtain an upper limit for CO, we assume the potential CO feature to be optically thin with an amplitude less than the S/N of the observed spectrum.

\section{Comparison to SN 1987A and limits on mixing during the explosion}
\label{sec:1987Acomparison}

In SN~2022acko, the C/O-rich layers are not mixed into the H-rich layers expanding with $\approx 2000-2500$~\kms\ as in SN~1987A.
At the epoch with similar expansion at the photosphere and light curve phase determined by the
H-recombination, SN~1987A showed extended mixing of H down to 800 km~s$^{-1}$ and $CO$ out to 3000~\kms \citep{Danziger1988,Hoeflich_1988,Meikle1988}, whereas SN~2022acko
   shows no evidence for any mixing during the explosion. 

From Table~\ref{tab:mir_lines} we see that the hydrogen lines in
SN~2022acko are in a 
velocity range from $\sim 2000-5000$~\kms.
The line profiles in the infared
exhibit the  traditional  P~Cygni shape. In the case of SN~1987A the
profiles, while still generally P~Cygni, showed a double-peaked
emission which is very apparent in   $Pa_\alpha$ as well as  the Balmer lines 
\citep{1987ESOC...26..147L,1987MNRAS.229P..15C}. 
 This was interpreted to be caused by mixing of
C/O into the H-rich envelope as well as of H into the C/O
core \citep{Hoeflich_1988}.
This would be from Rayleigh Taylor instabilities
caused by the shock propagation through the ejecta prior to the onset
of homology 
\citep{1989A&A...220..167M,Fryxell_etal_1991,1998ApJ...495..413N}.
Given  the lack of evidence for asymmetry in the H
lines, as well as a lack of $CO$ formation, we conclude that mixing in
SN~2022acko is significantly below that of SN~1987A, likely due to the
the difference in progenitor history.
Hence, any freshly formed $CO$ observable at later times in SN~2022acko should
 be attributed to $CO$ formation in the C/O core. 

\subsection{Molecules in the Circumstellar Material}
If $CO$ exists in the CSM it would add opacity due to cold $CO$ in the
surroundings of SN~2022acko. Since we do not see $CO$, any
pre-existing $CO$ is optically thin. The emission will be $\propto T$ in
the fundamental band with most of the power emitted in the first overtone
of the fundamental band. 
 Using the scaling in Fig.~\ref{fig:CO_fit}, the upper limits of $M(CO)$
 in the CSM at $T=1000$, 500, and 100~K are $10^{-8}$, $2 \times 10^{-8}
 $ and $10^{-7} M_\odot $, respectively.  If $CO$ is observed in
 SN~2022acko at later times, it will be due to freshly synthesized molecules.

\subsection{Pre-existing dust in the Interstellar and  Circumstellar Medium}
In principle, an upper limit for dust in the ISM and CSM can be
obtained the same way as for $CO$ above. However, the absorption to
emission ratio depends sensitively upon the grain-size
distribution. For spherical dust, the surface to mass ratio $\propto
1/R$ where $R$ is  the dust radius for particles larger than the
wavelength of light or is given by Rayleigh scattering for shorter
wavelengths. In the absence of observed dust signatures, we 
conclude that there is no evidence for ISM or CSM dust or a warm dust
component.

\clearpage

\section{Conclusions}
\label{sec:conclusions}

In this work, we presented the first \textit{JWST} spectrum of 
a CCSN, 2022acko. 
The data were  obtained  $\sim$52~days after the explosion epoch, overlapping with the H-recombination phase. 
Hydrogen lines mainly dominate the spectra of SN~2022acko but also contain spectral lines from C, N, Na, O, Mg, Br, Sc, and Fe. 
Combining our \textit{JWST} data with ground-based optical and NIR spectra, 
we created a full SED ranging from 0.4-25~$\mu$m.  The spectrum of \acko\ is found to resemble similar epoch spectra of SN~1987A and
SN~2005af. However, by these phases, SN~2005af already showed
forbidden lines while \acko\ does not. This may suggest that
SN~2022acko is evolving more slowly than SN~2005af. One should note
that the exact phase of SN~2005af is rather uncertain.

We have shown that the \textit{JWST} spectra provide new constraints
on the underlying physics  of SNe~IIP. Our analysis of the
multi-wavelength spectra allows us to  identify lines of  carbon,
nitrogen, and oxygen as well as 
s-process elements, which confirms the expected evolutionary
phases of a pre-SN red supergiant including AGB-like burning
conditions occurring in the ejecta leading to s-process production. 

We also demonstrate that placing limits on the flux level of the fundamental band of $CO$ can be used to measure the processes happening during the stellar evolution, the explosion, and within the environment. 
We see no extended mixing of the H and the C/O core is present at this epoch, and place tight upper  limits on the  $CO$ mass. We can rule out  significant  mixing via
 Raleigh Taylor instabilities in the ejecta or a heavily asymmetric
 explosion. Finally, we demonstrate that \textit{JWST} is needed for
 finding the pre-existing $CO$ mass, and conclude that  SN~2022acko
 has no detectable, pre-existing $CO$ ($< 10^{-8} M_\odot$), as well as
 no dust in the CSM. This important finding provides a clean slate for
 future \textit{JWST} observations which will probe molecule and dust
 formation in SN~2022acko at later times.  Any molecules or dust seen
 in future JWST epochs will be newly formed, since this epoch has
 served to rule out the presence of any pre-existing dust or molecules.

The observations from this work set a critical baseline for molecule and dust formation in SN~2022acko. Future JWST spectroscopic
observations of SN 2022acko will provide important information on the
formation of $CO$, $SiO$, and dust. 
Comparison of the rate of molecule and
dust formation with SNe 1987A, 2004dj, and 2005af will provide important clues
into the dust budget of SNe IIP in general as well as the variation
within the SNe IIP subclass.


\begin{acknowledgements}
We thank Jozsef Vinko for providing us with electronic versions of
Spitzer spectra.
MS acknowledges support by NASA grant JWST-GO-04436, JWST-GO-04217, JWST-GO-01860, JWST-GO-02666.
This work was supported by a NASA Keck PI Data Award, administered by the NASA Exoplanet Science Institute. The Observatory was made possible by the generous financial support of the W. M. Keck Foundation. The authors wish to recognize and acknowledge the very significant cultural role and reverence that the summit of Maunakea has always had within the indigenous Hawaiian community. We are most fortunate to have the opportunity to conduct observations from this mountain.
CA, PH. EB, and JD acknowledge support by NASA grant JWST-GO-02114,  JWST-GO-02122,  JWST-GO-04436, and  JWST-GO-04522.
Support for program \#2122 was provided by NASA through a grant from the Space Telescope Science  Institute, which is operated by the Association of Universities for Research in Astronomy, Inc., under NASA contract NAS 5-03127.
PH acknowledges support by the National Science Foundation (NSF) through grant AST-2306395 and NASA grant 80NSSC20K0538.
EB acknowledges support by NASA grant 80NSSC20K0538. 
MDS is funded by the Independent Research Fund Denmark (IRFD, grant number  10.46540/2032-00022B).
This publication was made possible through the support of an LSSTC Catalyst Fellowship to KAB funded through Grant 62192 from the John Templeton Foundation to LSST Corporation. The opinions expressed in this publication are those of the author(s) and do not necessarily reflect the views of LSSTC or the John Templeton Foundation.
ID acknowledges partial support by the Spanish project PID2021-123110NB 100 financed by MCIN/AEI/10.13039/501100011033/FEDER/UE. 
LG acknowledges financial support from the Spanish Ministerio de Ciencia e Innovaci\'on (MCIN), the Agencia Estatal de Investigaci\'on (AEI) 10.13039/501100011033, and the European Social Fund (ESF) "Investing in your future" under the 2019 Ram\'on y Cajal program RYC2019-027683-I and the PID2020-115253GA-I00 HOSTFLOWS project, from Centro Superior de Investigaciones Cient\'ificas (CSIC) under the PIE project 20215AT016, and the program Unidad de Excelencia Mar\'ia de Maeztu CEX2020-001058-M.
The research of YY is supported through a Bengier-Winslow-Robertson Fellowship.
SM acknowledges support from the Magnus Ehrnrooth Foundation and the Vilho, Yrj\"{o} and Kalle V\"{a}is\"{a}l\"{a} Foundation.
This work is based on observations made with the NASA/ESA/CSA James Webb Space Telescope. 
The data were obtained from the Mikulski Archive for Space Telescopes at the Space Telescope Science Institute, which is operated by the Association of Universities for Research in Astronomy, Inc., under NASA contract NAS 5-03127 for JWST. These observations are associated with program \#2122.

\clearpage

We acknowledge the Keck Infrared Transient Survey (KITS), which was executed primarily by members of the UC Santa Cruz transients team, who were supported in part by NASA grants NNG17PX03C, 80NSSC21K2076z, 80NSSC22K1513, 80NSSC22K1518; NSF grant AST–1911206; and by fellowships from the Alfred P. Sloan Foundation and the David and Lucile Packard Foundation to R.J.F. KITS was directly supported by NASA grant 80NSSC23K0301. C.D.K. is partly supported by a CIERA postdoctoral fellowship.
C.L. acknowledges support from the National Science Foundation Graduate Research Fellowship under grant No. DGE-2233066.
Y.-Z. Cai is supported by the National Natural Science Foundation of China (NSFC, Grant No. 12303054) and the International Centre of Supernovae, Yunnan Key Laboratory (No. 202302AN360001).
\end{acknowledgements}

\facilities{JWST (MRS/MIRI,NIRSpec/MIRI), MAST (JWST)}

\software{jwst (\citealp[ver. 1.9.4,][]{Bushouse2023_JWSTpipeline}), jdaviz (\citealp[ver. 3.2.1,][]{jdaviz}), HYDRA \citep{Hoeflich2003,Hoeflich2009,Hoeflich_etal_2017}, OpenDx (an open-sourced visualization package developed by IBM), Astropy \citep{astropy:2013, astropy:2018, astropy:2022}, NumPy \citep{numpy2020}, SciPy \citep{SciPy2020}, Matplotlib \citep{matplotlib}, pypeit \citep{2020JOSS....5.2308P}. }

\clearpage
\bibliographystyle{aasjournal}
\bibliography{refs}

\begin{thebibliography}{}
\expandafter\ifx\csname natexlab\endcsname\relax\def\natexlab#1{#1}\fi
\providecommand{\url}[1]{\href{#1}{#1}}
\providecommand{\dodoi}[1]{doi:~\href{http://doi.org/#1}{\nolinkurl{#1}}}
\providecommand{\doeprint}[1]{\href{http://ascl.net/#1}{\nolinkurl{http://ascl.net/#1}}}
\providecommand{\doarXiv}[1]{\href{https://arxiv.org/abs/#1}{\nolinkurl{https://arxiv.org/abs/#1}}}

\bibitem[{{Aitken} {et~al.}(1988){Aitken}, {Smith}, {James}, {Roche}, {Hyland}, \& {McGregor}}]{Aitken88}
{Aitken}, D.~K., {Smith}, C.~H., {James}, S.~D., {et~al.} 1988, \mnras, 235, 19P, \dodoi{10.1093/mnras/235.1.19P}

\bibitem[{{Anand} {et~al.}(2021){Anand}, {Lee}, {Van Dyk}, {Leroy}, {Rosolowsky}, {Schinnerer}, {Larson}, {Kourkchi}, {Kreckel}, {Scheuermann}, {Rizzi}, {Thilker}, {Tully}, {Bigiel}, {Blanc}, {Boquien}, {Chandar}, {Dale}, {Emsellem}, {Deger}, {Glover}, {Grasha}, {Groves}, {S. Klessen}, {Kruijssen}, {Querejeta}, {S{\'a}nchez-Bl{\'a}zquez}, {Schruba}, {Turner}, {Ubeda}, {Williams}, \& {Whitmore}}]{2021MNRAS.501.3621A}
{Anand}, G.~S., {Lee}, J.~C., {Van Dyk}, S.~D., {et~al.} 2021, \mnras, 501, 3621, \dodoi{10.1093/mnras/staa3668}

\bibitem[{{Astropy Collaboration} {et~al.}(2013){Astropy Collaboration}, {Robitaille}, {Tollerud}, {Greenfield}, {Droettboom}, {Bray}, {Aldcroft}, {Davis}, {Ginsburg}, {Price-Whelan}, {Kerzendorf}, {Conley}, {Crighton}, {Barbary}, {Muna}, {Ferguson}, {Grollier}, {Parikh}, {Nair}, {Unther}, {Deil}, {Woillez}, {Conseil}, {Kramer}, {Turner}, {Singer}, {Fox}, {Weaver}, {Zabalza}, {Edwards}, {Azalee Bostroem}, {Burke}, {Casey}, {Crawford}, {Dencheva}, {Ely}, {Jenness}, {Labrie}, {Lim}, {Pierfederici}, {Pontzen}, {Ptak}, {Refsdal}, {Servillat}, \& {Streicher}}]{astropy:2013}
{Astropy Collaboration}, {Robitaille}, T.~P., {Tollerud}, E.~J., {et~al.} 2013, \aap, 558, A33, \dodoi{10.1051/0004-6361/201322068}

\bibitem[{{Astropy Collaboration} {et~al.}(2018){Astropy Collaboration}, {Price-Whelan}, {Sip{\H{o}}cz}, {G{\"u}nther}, {Lim}, {Crawford}, {Conseil}, {Shupe}, {Craig}, {Dencheva}, {Ginsburg}, {Vand erPlas}, {Bradley}, {P{\'e}rez-Su{\'a}rez}, {de Val-Borro}, {Aldcroft}, {Cruz}, {Robitaille}, {Tollerud}, {Ardelean}, {Babej}, {Bach}, {Bachetti}, {Bakanov}, {Bamford}, {Barentsen}, {Barmby}, {Baumbach}, {Berry}, {Biscani}, {Boquien}, {Bostroem}, {Bouma}, {Brammer}, {Bray}, {Breytenbach}, {Buddelmeijer}, {Burke}, {Calderone}, {Cano Rodr{\'\i}guez}, {Cara}, {Cardoso}, {Cheedella}, {Copin}, {Corrales}, {Crichton}, {D'Avella}, {Deil}, {Depagne}, {Dietrich}, {Donath}, {Droettboom}, {Earl}, {Erben}, {Fabbro}, {Ferreira}, {Finethy}, {Fox}, {Garrison}, {Gibbons}, {Goldstein}, {Gommers}, {Greco}, {Greenfield}, {Groener}, {Grollier}, {Hagen}, {Hirst}, {Homeier}, {Horton}, {Hosseinzadeh}, {Hu}, {Hunkeler}, {Ivezi{\'c}}, {Jain}, {Jenness}, {Kanarek}, {Kendrew}, {Kern}, {Kerzendorf}, {Khvalko}, {King}, {Kirkby}, {Kulkarni},
  {Kumar}, {Lee}, {Lenz}, {Littlefair}, {Ma}, {Macleod}, {Mastropietro}, {McCully}, {Montagnac}, {Morris}, {Mueller}, {Mumford}, {Muna}, {Murphy}, {Nelson}, {Nguyen}, {Ninan}, {N{\"o}the}, {Ogaz}, {Oh}, {Parejko}, {Parley}, {Pascual}, {Patil}, {Patil}, {Plunkett}, {Prochaska}, {Rastogi}, {Reddy Janga}, {Sabater}, {Sakurikar}, {Seifert}, {Sherbert}, {Sherwood-Taylor}, {Shih}, {Sick}, {Silbiger}, {Singanamalla}, {Singer}, {Sladen}, {Sooley}, {Sornarajah}, {Streicher}, {Teuben}, {Thomas}, {Tremblay}, {Turner}, {Terr{\'o}n}, {van Kerkwijk}, {de la Vega}, {Watkins}, {Weaver}, {Whitmore}, {Woillez}, {Zabalza}, \& {Astropy Contributors}}]{astropy:2018}
{Astropy Collaboration}, {Price-Whelan}, A.~M., {Sip{\H{o}}cz}, B.~M., {et~al.} 2018, \aj, 156, 123, \dodoi{10.3847/1538-3881/aabc4f}

\bibitem[{{Astropy Collaboration} {et~al.}(2022){Astropy Collaboration}, {Price-Whelan}, {Lim}, {Earl}, {Starkman}, {Bradley}, {Shupe}, {Patil}, {Corrales}, {Brasseur}, {N{"o}the}, {Donath}, {Tollerud}, {Morris}, {Ginsburg}, {Vaher}, {Weaver}, {Tocknell}, {Jamieson}, {van Kerkwijk}, {Robitaille}, {Merry}, {Bachetti}, {G{"u}nther}, {Aldcroft}, {Alvarado-Montes}, {Archibald}, {B{'o}di}, {Bapat}, {Barentsen}, {Baz{'a}n}, {Biswas}, {Boquien}, {Burke}, {Cara}, {Cara}, {Conroy}, {Conseil}, {Craig}, {Cross}, {Cruz}, {D'Eugenio}, {Dencheva}, {Devillepoix}, {Dietrich}, {Eigenbrot}, {Erben}, {Ferreira}, {Foreman-Mackey}, {Fox}, {Freij}, {Garg}, {Geda}, {Glattly}, {Gondhalekar}, {Gordon}, {Grant}, {Greenfield}, {Groener}, {Guest}, {Gurovich}, {Handberg}, {Hart}, {Hatfield-Dodds}, {Homeier}, {Hosseinzadeh}, {Jenness}, {Jones}, {Joseph}, {Kalmbach}, {Karamehmetoglu}, {Ka{l}uszy{'n}ski}, {Kelley}, {Kern}, {Kerzendorf}, {Koch}, {Kulumani}, {Lee}, {Ly}, {Ma}, {MacBride}, {Maljaars}, {Muna}, {Murphy}, {Norman}, {O'Steen},
  {Oman}, {Pacifici}, {Pascual}, {Pascual-Granado}, {Patil}, {Perren}, {Pickering}, {Rastogi}, {Roulston}, {Ryan}, {Rykoff}, {Sabater}, {Sakurikar}, {Salgado}, {Sanghi}, {Saunders}, {Savchenko}, {Schwardt}, {Seifert-Eckert}, {Shih}, {Jain}, {Shukla}, {Sick}, {Simpson}, {Singanamalla}, {Singer}, {Singhal}, {Sinha}, {Sip{H{o}}cz}, {Spitler}, {Stansby}, {Streicher}, {{ {S}}umak}, {Swinbank}, {Taranu}, {Tewary}, {Tremblay}, {Val-Borro}, {Van Kooten}, {Vasovi{'c}}, {Verma}, {de Miranda Cardoso}, {Williams}, {Wilson}, {Winkel}, {Wood-Vasey}, {Xue}, {Yoachim}, {Zhang}, {Zonca}, \& {Astropy Project Contributors}}]{astropy:2022}
{Astropy Collaboration}, {Price-Whelan}, A.~M., {Lim}, P.~L., {et~al.} 2022, apj, 935, 167, \dodoi{10.3847/1538-4357/ac7c74}

\bibitem[{{Baron} {et~al.}(2003){Baron}, {Nugent}, {Branch}, {Hauschildt}, {Turatto}, \& {Cappellaro}}]{Baron03}
{Baron}, E., {Nugent}, P.~E., {Branch}, D., {et~al.} 2003, \apj, 586, 1199, \dodoi{10.1086/367888}

\bibitem[{{Bertoldi} {et~al.}(2003){Bertoldi}, {Cox}, {Neri}, {Carilli}, {Walter}, {Omont}, {Beelen}, {Henkel}, {Fan}, {Strauss}, \& {Menten}}]{Bertoldi2003}
{Bertoldi}, F., {Cox}, P., {Neri}, R., {et~al.} 2003, \aap, 409, L47, \dodoi{10.1051/0004-6361:20031345}

\bibitem[{{Bostroem} {et~al.}(2023){Bostroem}, {Dessart}, {Hillier}, {Lundquist}, {Andrews}, {Sand}, {Dong}, {Valenti}, {Haislip}, {Hoang}, {Hosseinzadeh}, {Janzen}, {Jencson}, {Jha}, {Kouprianov}, {Pearson}, {Meza Retamal}, {Reichart}, {Shrestha}, {Ashall}, {Baron}, {Brown}, {DerKacy}, {Farah}, {Galbany}, {Hern{\'a}ndez}, {Green}, {Hoeflich}, {Howell}, {Kwok}, {McCully}, {M{\"u}ller-Bravo}, {Newsome}, {Gonzalez}, {Pellegrino}, {Rho}, {Rowe}, {Schwab}, {Shahbandeh}, {Smith}, {Strader}, {Terreran}, {Van Dyk}, \& {Wyatt}}]{Bostroem23}
{Bostroem}, K.~A., {Dessart}, L., {Hillier}, D.~J., {et~al.} 2023, \apjl, 953, L18, \dodoi{10.3847/2041-8213/ace31c}

\bibitem[{{Bouchet} {et~al.}(1989){Bouchet}, {Moneti}, {Slezak}, {Le Bertre}, \& {Manfroid}}]{Bouchet89}
{Bouchet}, P., {Moneti}, A., {Slezak}, E., {Le Bertre}, T., \& {Manfroid}, J. 1989, \aaps, 80, 379

\bibitem[{{Bouchet} {et~al.}(2015){Bouchet}, {Garc{\'\i}a-Mar{\'\i}n}, {Lagage}, {Amiaux}, {Augu{\'e}res}, {Bauwens}, {Blommaert}, {Chen}, {Detre}, {Dicken}, {Dubreuil}, {Galdemard}, {Gastaud}, {Glasse}, {Gordon}, {Gougnaud}, {Guillard}, {Justtanont}, {Krause}, {Leboeuf}, {Longval}, {Martin}, {Mazy}, {Moreau}, {Olofsson}, {Ray}, {Rees}, {Renotte}, {Ressler}, {Ronayette}, {Salasca}, {Scheithauer}, {Sykes}, {Thelen}, {Wells}, {Wright}, \& {Wright}}]{Bouchet_2015}
{Bouchet}, P., {Garc{\'\i}a-Mar{\'\i}n}, M., {Lagage}, P.~O., {et~al.} 2015, \pasp, 127, 612, \dodoi{10.1086/682254}

\bibitem[{Bushouse {et~al.}(2023)Bushouse, Eisenhamer, Dencheva, Davies, Greenfield, Morrison, Hodge, Simon, Grumm, Droettboom, Slavich, Sosey, Pauly, Miller, Jedrzejewski, Hack, Davis, Crawford, Law, Gordon, Regan, Cara, MacDonald, Bradley, Shanahan, Jamieson, Teodoro, \& Williams}]{Bushouse2023_JWSTpipeline}
Bushouse, H., Eisenhamer, J., Dencheva, N., {et~al.} 2023, JWST Calibration Pipeline, 1.9.4,  Zenodo, \dodoi{10.5281/zenodo.7577320}

\bibitem[{{Catchpole} {et~al.}(1987){Catchpole}, {Menzies}, {Monk}, {Wargau}, {Pollaco}, {Carter}, {Whitelock}, {Marang}, {Laney}, {Balona}, {Feast}, {Lloyd Evans}, {Sekiguchi}, {Laing}, {Kilkenny}, {Spencer Jones}, {Roberts}, {Cousins}, {van Vuuren}, \& {Winkler}}]{1987MNRAS.229P..15C}
{Catchpole}, R.~M., {Menzies}, J.~W., {Monk}, A.~S., {et~al.} 1987, \mnras, 229, 15P, \dodoi{10.1093/mnras/229.1.15P}

\bibitem[{{Catchpole} {et~al.}(1988){Catchpole}, {Whitelock}, {Feast}, {Menzies}, {Glass}, {Marang}, {Laing}, {Spencer Jones}, {Roberts}, {Balona}, {Carter}, {Laney}, {Evans}, {Sekiguchi}, {Hutchinson}, {Maddison}, {Albinson}, {Evans}, {Allen}, {Winkler}, {Fairall}, {Corbally}, {Davies}, \& {Parker}}]{Catchpole1988}
{Catchpole}, R.~M., {Whitelock}, P.~A., {Feast}, M.~W., {et~al.} 1988, \mnras, 231, 75P, \dodoi{10.1093/mnras/231.1.75P}

\bibitem[{{Cernuschi} {et~al.}(1967){Cernuschi}, {Marsicano}, \& {Codina}}]{Cernuschi_1967}
{Cernuschi}, F., {Marsicano}, F., \& {Codina}, S. 1967, Annales d'Astrophysique, 30, 1039

\bibitem[{{Cherchneff} \& {Dwek}(2009)}]{Cherchneff_2009}
{Cherchneff}, I., \& {Dwek}, E. 2009, \apj, 703, 642, \dodoi{10.1088/0004-637X/703/1/642}

\bibitem[{{Chieffi} {et~al.}(2003){Chieffi}, {Dom{\'\i}nguez}, {H{\"o}flich}, {Limongi}, \& {Straniero}}]{2003MNRAS.345..111C}
{Chieffi}, A., {Dom{\'\i}nguez}, I., {H{\"o}flich}, P., {Limongi}, M., \& {Straniero}, O. 2003, \mnras, 345, 111, \dodoi{10.1046/j.1365-8711.2003.06958.x}

\bibitem[{{Danziger}(1988)}]{Danziger1988}
{Danziger}, I.~J. 1988, in Origin and Distribution of the Elements, ed. G.~J. {Mathews}, 407

\bibitem[{{Davis} {et~al.}(2019){Davis}, {Hsiao}, {Ashall}, {Hoeflich}, {Phillips}, {Marion}, {Kirshner}, {Morrell}, {Sand}, {Burns}, {Contreras}, {Stritzinger}, {Anderson}, {Baron}, {Diamond}, {Guti{\'e}rrez}, {Hamuy}, {Holmbo}, {Kasliwal}, {Krisciunas}, {Kumar}, {Lu}, {Pessi}, {Piro}, {Prieto}, {Shahbandeh}, \& {Suntzeff}}]{Davis_2019}
{Davis}, S., {Hsiao}, E.~Y., {Ashall}, C., {et~al.} 2019, \apj, 887, 4, \dodoi{10.3847/1538-4357/ab4c40}

\bibitem[{{de Vaucouleurs} {et~al.}(1991){de Vaucouleurs}, {de Vaucouleurs}, {Corwin}, {Buta}, {Paturel}, \& {Fouque}}]{1991rc3..book.....D}
{de Vaucouleurs}, G., {de Vaucouleurs}, A., {Corwin}, Herold~G., J., {et~al.} 1991, {Third Reference Catalogue of Bright Galaxies}

\bibitem[{{Dessart} {et~al.}(2013){Dessart}, {Hillier}, {Waldman}, \& {Livne}}]{2013MNRAS.433.1745D}
{Dessart}, L., {Hillier}, D.~J., {Waldman}, R., \& {Livne}, E. 2013, \mnras, 433, 1745, \dodoi{10.1093/mnras/stt861}

\bibitem[{Developers {et~al.}(2023)Developers, Averbukh, Bradley, Buikhuizen, Busko, Cherinka, Conroy, Earl, Fox, Geda, Jones, Kotler, Lim, Morris, Nguyen, O'Steen, Ogaz, Ogle, Otor, Pacifici, Robitaille, Tollerud, Volfman, \& Karatay}]{jdaviz}
Developers, J., Averbukh, J., Bradley, L., {et~al.} 2023, Jdaviz, 3.2.1,  Zenodo, \dodoi{10.5281/zenodo.7600492}

\bibitem[{{Duschinger} {et~al.}(1995){Duschinger}, {Puls}, {Branch}, {Hoeflich}, \& {Gabler}}]{Duschinger_etal_1995}
{Duschinger}, M., {Puls}, J., {Branch}, D., {Hoeflich}, P., \& {Gabler}, A. 1995, \aap, 297, 802

\bibitem[{{Dwek} {et~al.}(2007){Dwek}, {Galliano}, \& {Jones}}]{Dwek_2007}
{Dwek}, E., {Galliano}, F., \& {Jones}, A.~P. 2007, \apj, 662, 927, \dodoi{10.1086/518430}

\bibitem[{{Ferrarotti} \& {Gail}(2006)}]{Ferrarotti2006}
{Ferrarotti}, A.~S., \& {Gail}, H.~P. 2006, \aap, 447, 553, \dodoi{10.1051/0004-6361:20041198}

\bibitem[{{Fox} {et~al.}(2010){Fox}, {Chevalier}, {Dwek}, {Skrutskie}, {Sugerman}, \& {Leisenring}}]{Fox10}
{Fox}, O.~D., {Chevalier}, R.~A., {Dwek}, E., {et~al.} 2010, \apj, 725, 1768, \dodoi{10.1088/0004-637X/725/2/1768}

\bibitem[{{Fryxell} {et~al.}(1991){Fryxell}, {Mueller}, \& {Arnett}}]{Fryxell_etal_1991}
{Fryxell}, B., {Mueller}, E., \& {Arnett}, D. 1991, \apj, 367, 619, \dodoi{10.1086/169657}

\bibitem[{{Gall} {et~al.}(2011){Gall}, {Hjorth}, \& {Andersen}}]{Gall_2011_a}
{Gall}, C., {Hjorth}, J., \& {Andersen}, A.~C. 2011, \aapr, 19, 43, \dodoi{10.1007/s00159-011-0043-7}

\bibitem[{{G{\'e}rard} {et~al.}(2000){G{\'e}rard}, {Hubert}, {Bisikalo}, \& {Shematovich}}]{Gerard_etal_2000}
{G{\'e}rard}, J.-C., {Hubert}, B., {Bisikalo}, D.~V., \& {Shematovich}, V.~I. 2000, \jgr, 105, 15795, \dodoi{10.1029/1999JA002002}

\bibitem[{{Gerardy} {et~al.}(2000){Gerardy}, {Fesen}, {H{\"o}flich}, \& {Wheeler}}]{Gerardy_etal_2000}
{Gerardy}, C.~L., {Fesen}, R.~A., {H{\"o}flich}, P., \& {Wheeler}, J.~C. 2000, \aj, 119, 2968, \dodoi{10.1086/301390}

\bibitem[{{Gerardy} {et~al.}(2002){Gerardy}, {Fesen}, {Nomoto}, {Maeda}, {Hoflich}, \& {Wheeler}}]{Gerardy_etal_2002}
{Gerardy}, C.~L., {Fesen}, R.~A., {Nomoto}, K., {et~al.} 2002, \pasj, 54, 905, \dodoi{10.1093/pasj/54.6.905}

\bibitem[{{Gonz{\'a}lez Delgado} {et~al.}(2003){Gonz{\'a}lez Delgado}, {Olofsson}, {Kerschbaum}, {Sch{\"o}ier}, {Lindqvist}, \& {Groenewegen}}]{Gonzo2003}
{Gonz{\'a}lez Delgado}, D., {Olofsson}, H., {Kerschbaum}, F., {et~al.} 2003, \aap, 411, 123, \dodoi{10.1051/0004-6361:20031068}

\bibitem[{{Harris} {et~al.}(2020){Harris}, {Millman}, {van der Walt}, {Gommers}, {Virtanen}, {Cournapeau}, {Wieser}, {Taylor}, {Berg}, {Smith}, {Kern}, {Picus}, {Hoyer}, {van Kerkwijk}, {Brett}, {Haldane}, {del R{\'\i}o}, {Wiebe}, {Peterson}, {G{\'e}rard-Marchant}, {Sheppard}, {Reddy}, {Weckesser}, {Abbasi}, {Gohlke}, \& {Oliphant}}]{numpy2020}
{Harris}, C.~R., {Millman}, K.~J., {van der Walt}, S.~J., {et~al.} 2020, \nat, 585, 357, \dodoi{10.1038/s41586-020-2649-2}

\bibitem[{{Hoeflich}(1988)}]{Hoeflich_1988}
{Hoeflich}, P. 1988, \pasa, 7, 434, \dodoi{10.1017/S1323358000022608}

\bibitem[{{Hoeflich} {et~al.}(2017){Hoeflich}, {Hsiao}, {Ashall}, {Burns}, {Diamond}, {Phillips}, {Sand}, {Stritzinger}, {Suntzeff}, {Contreras}, {Krisciunas}, {Morrell}, \& {Wang}}]{Hoeflich_etal_2017}
{Hoeflich}, P., {Hsiao}, E.~Y., {Ashall}, C., {et~al.} 2017, \apj, 846, 58, \dodoi{10.3847/1538-4357/aa84b2}

\bibitem[{{H{\"o}flich}(1990)}]{Hoeflich_1990_Habil}
{H{\"o}flich}, P. 1990, PhD thesis, -

\bibitem[{{H{\"o}flich}(2003)}]{Hoeflich2003}
{H{\"o}flich}, P. 2003, in Astronomical Society of the Pacific Conference Series, Vol. 288, Stellar Atmosphere Modeling, ed. I.~{Hubeny}, D.~{Mihalas}, \& K.~{Werner}, 185

\bibitem[{{H{\"o}flich}(2009)}]{Hoeflich2009}
{H{\"o}flich}, P. 2009, in American Institute of Physics Conference Series, Vol. 1171, Recent Directions in Astrophysical Quantitative Spectroscopy and Radiation Hydrodynamics, ed. I.~{Hubeny}, J.~M. {Stone}, K.~{MacGregor}, \& K.~{Werner}, 161--172, \dodoi{10.1063/1.3250057}

\bibitem[{{H{\"o}flich} {et~al.}(2002){H{\"o}flich}, {Gerardy}, {Fesen}, \& {Sakai}}]{Hoeflich_2002}
{H{\"o}flich}, P., {Gerardy}, C.~L., {Fesen}, R.~A., \& {Sakai}, S. 2002, \apj, 568, 791, \dodoi{10.1086/339063}

\bibitem[{{Hoyle} \& {Wickramasinghe}(1970)}]{Hoyle_1970}
{Hoyle}, F., \& {Wickramasinghe}, N.~C. 1970, \nat, 226, 62, \dodoi{10.1038/226062a0}

\bibitem[{{Hunter}(2007)}]{matplotlib}
{Hunter}, J.~D. 2007, Computing in Science and Engineering, 9, 90, \dodoi{10.1109/MCSE.2007.55}

\bibitem[{{Jakobsen} {et~al.}(2022){Jakobsen}, {Ferruit}, {Alves de Oliveira}, {Arribas}, {Bagnasco}, {Barho}, {Beck}, {Birkmann}, {B{\"o}ker}, {Bunker}, {Charlot}, {de Jong}, {de Marchi}, {Ehrenwinkler}, {Falcolini}, {Fels}, {Franx}, {Franz}, {Funke}, {Giardino}, {Gnata}, {Holota}, {Honnen}, {Jensen}, {Jentsch}, {Johnson}, {Jollet}, {Karl}, {Kling}, {K{\"o}hler}, {Kolm}, {Kumari}, {Lander}, {Lemke}, {L{\'o}pez-Caniego}, {L{\"u}tzgendorf}, {Maiolino}, {Manjavacas}, {Marston}, {Maschmann}, {Maurer}, {Messerschmidt}, {Moseley}, {Mosner}, {Mott}, {Muzerolle}, {Pirzkal}, {Pittet}, {Plitzke}, {Posselt}, {Rapp}, {Rauscher}, {Rawle}, {Rix}, {R{\"o}del}, {Rumler}, {Sabbi}, {Salvignol}, {Schmid}, {Sirianni}, {Smith}, {Strada}, {te Plate}, {Valenti}, {Wettemann}, {Wiehe}, {Wiesmayer}, {Willott}, {Wright}, {Zeidler}, \& {Zincke}}]{Jakobsen_2022}
{Jakobsen}, P., {Ferruit}, P., {Alves de Oliveira}, C., {et~al.} 2022, \aap, 661, A80, \dodoi{10.1051/0004-6361/202142663}

\bibitem[{{Kotak} {et~al.}(2005{\natexlab{a}}){Kotak}, {Meikle}, {van Dyk}, {H{\"o}flich}, \& {Mattila}}]{Kotak2005}
{Kotak}, R., {Meikle}, P., {van Dyk}, S.~D., {H{\"o}flich}, P.~A., \& {Mattila}, S. 2005{\natexlab{a}}, \apjl, 628, L123, \dodoi{10.1086/432719}

\bibitem[{{Kotak} {et~al.}(2005{\natexlab{b}}){Kotak}, {Meikle}, {van Dyk}, {H{\"o}flich}, \& {Mattila}}]{Kotak_etal_2005_04dj}
---. 2005{\natexlab{b}}, \apjl, 628, L123, \dodoi{10.1086/432719}

\bibitem[{{Kotak} {et~al.}(2006){Kotak}, {Meikle}, {Pozzo}, {van Dyk}, {Farrah}, {Fesen}, {Filippenko}, {Foley}, {Fransson}, {Gerardy}, {H{\"o}flich}, {Lundqvist}, {Mattila}, {Sollerman}, \& {Wheeler}}]{Kotak_etal_2006_05af}
{Kotak}, R., {Meikle}, P., {Pozzo}, M., {et~al.} 2006, \apjl, 651, L117, \dodoi{10.1086/509655}

\bibitem[{{Kotak} {et~al.}(2009){Kotak}, {Meikle}, {Farrah}, {Gerardy}, {Foley}, {Van Dyk}, {Fransson}, {Lundqvist}, {Sollerman}, {Fesen}, {Filippenko}, {Mattila}, {Silverman}, {Andersen}, {H{\"o}flich}, {Pozzo}, \& {Wheeler}}]{2009ApJ...704..306K}
{Kotak}, R., {Meikle}, W.~P.~S., {Farrah}, D., {et~al.} 2009, \apj, 704, 306, \dodoi{10.1088/0004-637X/704/1/306}

\bibitem[{{Larson} {et~al.}(1987){Larson}, {Drapatz}, {Mumma}, \& {Weaver}}]{1987ESOC...26..147L}
{Larson}, H.~P., {Drapatz}, S., {Mumma}, M.~J., \& {Weaver}, H.~A. 1987, in European Southern Observatory Conference and Workshop Proceedings, Vol.~26, European Southern Observatory Conference and Workshop Proceedings, 147

\bibitem[{{Lee} {et~al.}(2023){Lee}, {Sandstrom}, {Leroy}, {Thilker}, {Schinnerer}, {Rosolowsky}, {Larson}, {Egorov}, {Williams}, {Schmidt}, {Emsellem}, {Anand}, {Barnes}, {Belfiore}, {Be{\v{s}}li{\'c}}, {Bigiel}, {Blanc}, {Bolatto}, {Boquien}, {den Brok}, {Cao}, {Chandar}, {Chastenet}, {Chevance}, {Chiang}, {Congiu}, {Dale}, {Deger}, {Eibensteiner}, {Faesi}, {Glover}, {Grasha}, {Groves}, {Hassani}, {Henny}, {Henshaw}, {Hoyer}, {Hughes}, {Jeffreson}, {Jim{\'e}nez-Donaire}, {Kim}, {Kim}, {Klessen}, {Koch}, {Kreckel}, {Kruijssen}, {Li}, {Liu}, {Lopez}, {Maschmann}, {Chen}, {Meidt}, {Murphy}, {Neumann}, {Neumayer}, {Pan}, {Pessa}, {Pety}, {Querejeta}, {Pinna}, {Rodr{\'\i}guez}, {Saito}, {S{\'a}nchez-Bl{\'a}zquez}, {Santoro}, {Sardone}, {Smith}, {Sormani}, {Scheuermann}, {Stuber}, {Sutter}, {Sun}, {Teng}, {Tre{\ss}}, {Usero}, {Watkins}, {Whitmore}, \& {Razza}}]{Lee23}
{Lee}, J.~C., {Sandstrom}, K.~M., {Leroy}, A.~K., {et~al.} 2023, \apjl, 944, L17, \dodoi{10.3847/2041-8213/acaaae}

\bibitem[{{Lepp} {et~al.}(1990){Lepp}, {Dalgarno}, \& {McCray}}]{1990ApJ...358..262L}
{Lepp}, S., {Dalgarno}, A., \& {McCray}, R. 1990, \apj, 358, 262, \dodoi{10.1086/168981}

\bibitem[{{Li} {et~al.}(2022){Li}, {Cai}, {Zhai}, {Zhang}, \& {Wang}}]{Li22}
{Li}, L., {Cai}, Y., {Zhai}, Q., {Zhang}, J., \& {Wang}, X. 2022, Transient Name Server Classification Report, 2022-3549, 1

\bibitem[{{Li} {et~al.}(2020){Li}, {Wang}, {Fan}, {Wu}, {Jiang}, {Ba{\~n}ados}, {Venemans}, {Shao}, {Li}, {Zhang}, {Zhang}, {Wagg}, {Decarli}, {Mazzucchelli}, {Omont}, \& {Bertoldi}}]{Li2020}
{Li}, Q., {Wang}, R., {Fan}, X., {et~al.} 2020, \apj, 900, 12, \dodoi{10.3847/1538-4357/aba52d}

\bibitem[{{Liu} \& {Dalgarno}(1995)}]{Liu_Dalgarno_1995}
{Liu}, W., \& {Dalgarno}, A. 1995, \apj, 454, 472, \dodoi{10.1086/176498}

\bibitem[{{Lundquist} {et~al.}(2022){Lundquist}, {Pearson}, {Paraskeva}, {Hoang}, {Meza}, {Valenti}, {Shrestha}, {Sand}, {Wyatt}, {Andrews}, {Janzen}, {Bostroem}, {Jencson}, {Hosseinzadeh}, \& {Dong}}]{2022TNSTR3543....1L}
{Lundquist}, M., {Pearson}, J., {Paraskeva}, E., {et~al.} 2022, Transient Name Server Discovery Report, 2022-3543, 1

\bibitem[{{Maiolino} {et~al.}(2004){Maiolino}, {Schneider}, {Oliva}, {Bianchi}, {Ferrara}, {Mannucci}, {Pedani}, \& {Roca Sogorb}}]{Maiolino_2004}
{Maiolino}, R., {Schneider}, R., {Oliva}, E., {et~al.} 2004, \nat, 431, 533, \dodoi{10.1038/nature02930}

\bibitem[{{Mazzali} {et~al.}(1992){Mazzali}, {Lucy}, \& {Butler}}]{1992A&A...258..399M}
{Mazzali}, P.~A., {Lucy}, L.~B., \& {Butler}, K. 1992, \aap, 258, 399

\bibitem[{{Meikle}(1988)}]{Meikle1988}
{Meikle}, W.~P.~S. 1988, \pasa, 7, 473, \dodoi{10.1017/S1323358000022669}

\bibitem[{{Meikle} {et~al.}(1989){Meikle}, {Allen}, {Spyromilio}, \& {Varani}}]{Meikle1989}
{Meikle}, W.~P.~S., {Allen}, D.~A., {Spyromilio}, J., \& {Varani}, G.~F. 1989, \mnras, 238, 193, \dodoi{10.1093/mnras/238.1.193}

\bibitem[{{Meikle} {et~al.}(1993){Meikle}, {Spyromilio}, {Allen}, {Varani}, \& {Cumming}}]{1993MNRAS.261..535M}
{Meikle}, W.~P.~S., {Spyromilio}, J., {Allen}, D.~A., {Varani}, G.~F., \& {Cumming}, R.~J. 1993, \mnras, 261, 535, \dodoi{10.1093/mnras/261.3.535}

\bibitem[{{Meikle} {et~al.}(2007{\natexlab{a}}){Meikle}, {Mattila}, {Pastorello}, {Gerardy}, {Kotak}, {Sollerman}, {Van Dyk}, {Farrah}, {Filippenko}, {H{\"o}flich}, {Lundqvist}, {Pozzo}, \& {Wheeler}}]{2007ApJ...665..608M}
{Meikle}, W.~P.~S., {Mattila}, S., {Pastorello}, A., {et~al.} 2007{\natexlab{a}}, \apj, 665, 608, \dodoi{10.1086/519733}

\bibitem[{{Meikle} {et~al.}(2007{\natexlab{b}}){Meikle}, {Mattila}, {Pastorello}, {Gerardy}, {Kotak}, {Sollerman}, {Van Dyk}, {Farrah}, {Filippenko}, {H{\"o}flich}, {Lundqvist}, {Pozzo}, \& {Wheeler}}]{Meikle_etal_2007}
---. 2007{\natexlab{b}}, \apj, 665, 608, \dodoi{10.1086/519733}

\bibitem[{{Muller} {et~al.}(1989){Muller}, {Hillebrandt}, {Orio}, {Hoflich}, {Monchmeyer}, \& {Fryxell}}]{1989A&A...220..167M}
{Muller}, E., {Hillebrandt}, W., {Orio}, M., {et~al.} 1989, \aap, 220, 167

\bibitem[{{Nagataki} {et~al.}(1998){Nagataki}, {Shimizu}, \& {Sato}}]{1998ApJ...495..413N}
{Nagataki}, S., {Shimizu}, T.~M., \& {Sato}, K. 1998, \apj, 495, 413, \dodoi{10.1086/305258}

\bibitem[{{Nozawa} {et~al.}(2003){Nozawa}, {Kozasa}, {Umeda}, {Maeda}, \& {Nomoto}}]{Nozawa_2003}
{Nozawa}, T., {Kozasa}, T., {Umeda}, H., {Maeda}, K., \& {Nomoto}, K. 2003, \apj, 598, 785, \dodoi{10.1086/379011}

\bibitem[{{Nozawa} {et~al.}(2008){Nozawa}, {Kozasa}, {Tominaga}, {Sakon}, {Tanaka}, {Suzuki}, {Nomoto}, {Maeda}, {Umeda}, {Limongi}, \& {Onaka}}]{Nozawa_2008}
{Nozawa}, T., {Kozasa}, T., {Tominaga}, N., {et~al.} 2008, \apj, 684, 1343, \dodoi{10.1086/589961}

\bibitem[{{Perrin} {et~al.}(2014){Perrin}, {Sivaramakrishnan}, {Lajoie}, {Elliott}, {Pueyo}, {Ravindranath}, \& {Albert}}]{Perrin14}
{Perrin}, M.~D., {Sivaramakrishnan}, A., {Lajoie}, C.-P., {et~al.} 2014, in Society of Photo-Optical Instrumentation Engineers (SPIE) Conference Series, Vol. 9143, Space Telescopes and Instrumentation 2014: Optical, Infrared, and Millimeter Wave, ed. J.~{Oschmann}, Jacobus~M., M.~{Clampin}, G.~G. {Fazio}, \& H.~A. {MacEwen}, 91433X, \dodoi{10.1117/12.2056689}

\bibitem[{{Priddey} {et~al.}(2003){Priddey}, {Isaak}, {McMahon}, {Robson}, \& {Pearson}}]{Priddey2003}
{Priddey}, R.~S., {Isaak}, K.~G., {McMahon}, R.~G., {Robson}, E.~I., \& {Pearson}, C.~P. 2003, \mnras, 344, L74, \dodoi{10.1046/j.1365-8711.2003.07076.x}

\bibitem[{{Prochaska} {et~al.}(2020){Prochaska}, {Hennawi}, {Westfall}, {Cooke}, {Wang}, {Hsyu}, {Davies}, {Farina}, \& {Pelliccia}}]{2020JOSS....5.2308P}
{Prochaska}, J., {Hennawi}, J., {Westfall}, K., {et~al.} 2020, The Journal of Open Source Software, 5, 2308, \dodoi{10.21105/joss.02308}

\bibitem[{{Ressler} {et~al.}(2015){Ressler}, {Sukhatme}, {Franklin}, {Mahoney}, {Thelen}, {Bouchet}, {Colbert}, {Cracraft}, {Dicken}, {Gastaud}, {Goodson}, {Eccleston}, {Moreau}, {Rieke}, \& {Schneider}}]{Ressler_2015}
{Ressler}, M.~E., {Sukhatme}, K.~G., {Franklin}, B.~R., {et~al.} 2015, \pasp, 127, 675, \dodoi{10.1086/682258}

\bibitem[{{Rho} {et~al.}(2018){Rho}, {Geballe}, {Banerjee}, {Dessart}, {Evans}, \& {Joshi}}]{Rho_etal_2018}
{Rho}, J., {Geballe}, T.~R., {Banerjee}, D.~P.~K., {et~al.} 2018, \apjl, 864, L20, \dodoi{10.3847/2041-8213/aad77f}

\bibitem[{{Rho} {et~al.}(2019){Rho}, {Shahbandeh}, {Hsiao}, {Davis}, {Brown}, {Valenti}, {Bostroem}, {Hiramatsu}, {Burke}, {Howell}, {McCully}, {Szalai}, {Banerjee}, {Geballe}, {Dessart}, {Evans}, {Galbany}, \& {Team.}}]{Rho_2019_ATEL}
{Rho}, J., {Shahbandeh}, M., {Hsiao}, E., {et~al.} 2019, The Astronomer's Telegram, 12897, 1

\bibitem[{{Rho} {et~al.}(2021){Rho}, {Evans}, {Geballe}, {Banerjee}, {Hoeflich}, {Shahbandeh}, {Valenti}, {Yoon}, {Jin}, {Williamson}, {Modjaz}, {Hiramatsu}, {Howell}, {Pellegrino}, {Vink{\'o}}, {Cartier}, {Burke}, {McCully}, {An}, {Cha}, {Pritchard}, {Wang}, {Andrews}, {Galbany}, {Van Dyk}, {Graham}, {Blinnikov}, {Joshi}, {P{\'a}l}, {Kriskovics}, {Ordasi}, {Szakats}, {Vida}, {Chen}, {Li}, {Zhang}, \& {Yan}}]{Rho21}
{Rho}, J., {Evans}, A., {Geballe}, T.~R., {et~al.} 2021, \apj, 908, 232, \dodoi{10.3847/1538-4357/abd850}

\bibitem[{{Rieke} {et~al.}(2022){Rieke}, {Su}, {Sloan}, \& {Schlawin}}]{Rieke_2022}
{Rieke}, G.~H., {Su}, K., {Sloan}, G.~C., \& {Schlawin}, E. 2022, \aj, 163, 45, \dodoi{10.3847/1538-3881/ac3b5d}

\bibitem[{{Rieke} {et~al.}(2015){Rieke}, {Wright}, {B{\"o}ker}, {Bouwman}, {Colina}, {Glasse}, {Gordon}, {Greene}, {G{\"u}del}, {Henning}, {Justtanont}, {Lagage}, {Meixner}, {N{\o}rgaard-Nielsen}, {Ray}, {Ressler}, {van Dishoeck}, \& {Waelkens}}]{Rieke_2015}
{Rieke}, G.~H., {Wright}, G.~S., {B{\"o}ker}, T., {et~al.} 2015, \pasp, 127, 584, \dodoi{10.1086/682252}

\bibitem[{{Sarangi} \& {Cherchneff}(2015)}]{Sarangi_2015}
{Sarangi}, A., \& {Cherchneff}, I. 2015, \aap, 575, A95, \dodoi{10.1051/0004-6361/201424969}

\bibitem[{{Sarangi} {et~al.}(2018){Sarangi}, {Matsuura}, \& {Micelotta}}]{Sarangi_2018}
{Sarangi}, A., {Matsuura}, M., \& {Micelotta}, E.~R. 2018, \ssr, 214, 63, \dodoi{10.1007/s11214-018-0492-7}

\bibitem[{{Schneider} {et~al.}(2004){Schneider}, {Ferrara}, \& {Salvaterra}}]{Schneider_2004}
{Schneider}, R., {Ferrara}, A., \& {Salvaterra}, R. 2004, \mnras, 351, 1379, \dodoi{10.1111/j.1365-2966.2004.07876.x}

\bibitem[{{Shahbandeh} {et~al.}(2022){Shahbandeh}, {Hsiao}, {Ashall}, {Teffs}, {Hoeflich}, {Morrell}, {Phillips}, {Anderson}, {Baron}, {Burns}, {Contreras}, {Davis}, {Diamond}, {Folatelli}, {Galbany}, {Gall}, {Hachinger}, {Holmbo}, {Karamehmetoglu}, {Kasliwal}, {Kirshner}, {Krisciunas}, {Kumar}, {Lu}, {Marion}, {Mazzali}, {Piro}, {Sand}, {Stritzinger}, {Suntzeff}, {Taddia}, \& {Uddin}}]{Shahbandeh_2022}
{Shahbandeh}, M., {Hsiao}, E.~Y., {Ashall}, C., {et~al.} 2022, \apj, 925, 175, \dodoi{10.3847/1538-4357/ac4030}

\bibitem[{{Shahbandeh} {et~al.}(2023){Shahbandeh}, {Sarangi}, {Temim}, {Szalai}, {Fox}, {Tinyanont}, {Dwek}, {Dessart}, {Filippenko}, {Brink}, {Foley}, {Jencson}, {Pierel}, {Zs{\'\i}ros}, {Rest}, {Zheng}, {Andrews}, {Clayton}, {De}, {Engesser}, {Gezari}, {Gomez}, {Gonzaga}, {Johansson}, {Kasliwal}, {Lau}, {De Looze}, {Marston}, {Milisavljevic}, {O'Steen}, {Siebert}, {Skrutskie}, {Smith}, {Strolger}, {Van Dyk}, {Wang}, {Williams}, {Williams}, {Xiao}, \& {Yang}}]{Shahbandeh_2023}
{Shahbandeh}, M., {Sarangi}, A., {Temim}, T., {et~al.} 2023, \mnras, 523, 6048, \dodoi{10.1093/mnras/stad1681}

\bibitem[{{Sharp} \& {Hoeflich}(1990)}]{Sharp_1990}
{Sharp}, C.~M., \& {Hoeflich}, P. 1990, \apss, 171, 213, \dodoi{10.1007/BF00646849}

\bibitem[{{Sharp} \& {H{\"o}flich}(1989)}]{1989HiA.....8..207S}
{Sharp}, C.~M., \& {H{\"o}flich}, P. 1989, Highlights of Astronomy, 8, 207

\bibitem[{{Sluder} {et~al.}(2018){Sluder}, {Milosavljevi{\'c}}, \& {Montgomery}}]{Sluder_2018}
{Sluder}, A., {Milosavljevi{\'c}}, M., \& {Montgomery}, M.~H. 2018, \mnras, 480, 5580, \dodoi{10.1093/mnras/sty2060}

\bibitem[{{Spyromilio} {et~al.}(1988){Spyromilio}, {Meikle}, {Learner}, \& {Allen}}]{Spyromilio_1988}
{Spyromilio}, J., {Meikle}, W.~P.~S., {Learner}, R.~C.~M., \& {Allen}, D.~A. 1988, \nat, 334, 327, \dodoi{10.1038/334327a0}

\bibitem[{{Szalai} {et~al.}(2019){Szalai}, {Vink{\'o}}, {K{\"o}nyves-T{\'o}th}, {Nagy}, {Bostroem}, {S{\'a}rneczky}, {Brown}, {Pejcha}, {B{\'o}di}, {Cseh}, {Cs{\"o}rnyei}, {Dencs}, {Hanyecz}, {Ign{\'a}cz}, {Kalup}, {Kriskovics}, {Ordasi}, {P{\'a}l}, {Seli}, {S{\'o}dor}, {Szak{\'a}ts}, {Vida}, {Zsidi}, {Konkoly Team}, {Arcavi}, {Ashall}, {Burke}, {Galbany}, {Hiramatsu}, {Hosseinzadeh}, {Hsiao}, {Howell}, {McCully}, {Moran}, {Rho}, {Sand}, {Shahbandeh}, {Valenti}, {Wang}, {Wheeler}, \& {Supernova Project}}]{Szalai19}
{Szalai}, T., {Vink{\'o}}, J., {K{\"o}nyves-T{\'o}th}, R., {et~al.} 2019, \apj, 876, 19, \dodoi{10.3847/1538-4357/ab12d0}

\bibitem[{{Tartaglia} {et~al.}(2018){Tartaglia}, {Sand}, {Valenti}, {Wyatt}, {Anderson}, {Arcavi}, {Ashall}, {Botticella}, {Cartier}, {Chen}, {Cikota}, {Coulter}, {Della Valle}, {Foley}, {Gal-Yam}, {Galbany}, {Gall}, {Haislip}, {Harmanen}, {Hosseinzadeh}, {Howell}, {Hsiao}, {Inserra}, {Jha}, {Kankare}, {Kilpatrick}, {Kouprianov}, {Kuncarayakti}, {Maccarone}, {Maguire}, {Mattila}, {Mazzali}, {McCully}, {Melandri}, {Morrell}, {Phillips}, {Pignata}, {Piro}, {Prentice}, {Reichart}, {Rojas-Bravo}, {Smartt}, {Smith}, {Sollerman}, {Stritzinger}, {Sullivan}, {Taddia}, \& {Young}}]{DLT40}
{Tartaglia}, L., {Sand}, D.~J., {Valenti}, S., {et~al.} 2018, \apj, 853, 62, \dodoi{10.3847/1538-4357/aaa014}

\bibitem[{{Tinyanont} {et~al.}(2019){Tinyanont}, {Kasliwal}, {Krafton}, {Lau}, {Rho}, {Leonard}, {De}, {Jencson}, {Mawet}, {Millar-Blanchaer}, {Nilsson}, {Yan}, {Gehrz}, {Helou}, {Van Dyk}, {Serabyn}, {Fox}, \& {Clayton}}]{Tinyanont19}
{Tinyanont}, S., {Kasliwal}, M.~M., {Krafton}, K., {et~al.} 2019, \apj, 873, 127, \dodoi{10.3847/1538-4357/ab0897}

\bibitem[{{Todini} \& {Ferrara}(2001)}]{Todini_2001}
{Todini}, P., \& {Ferrara}, A. 2001, \mnras, 325, 726, \dodoi{10.1046/j.1365-8711.2001.04486.x}

\bibitem[{{Turatto} {et~al.}(1993){Turatto}, {Cappellaro}, {Benetti}, \& {Danziger}}]{1993MNRAS.265..471T}
{Turatto}, M., {Cappellaro}, E., {Benetti}, S., \& {Danziger}, I.~J. 1993, \mnras, 265, 471, \dodoi{10.1093/mnras/265.2.471}

\bibitem[{{Virtanen} {et~al.}(2020){Virtanen}, {Gommers}, {Oliphant}, {Haberland}, {Reddy}, {Cournapeau}, {Burovski}, {Peterson}, {Weckesser}, {Bright}, {van der Walt}, {Brett}, {Wilson}, {Millman}, {Mayorov}, {Nelson}, {Jones}, {Kern}, {Larson}, {Carey}, {Polat}, {Feng}, {Moore}, {VanderPlas}, {Laxalde}, {Perktold}, {Cimrman}, {Henriksen}, {Quintero}, {Harris}, {Archibald}, {Ribeiro}, {Pedregosa}, {van Mulbregt}, \& {SciPy 1. 0 Contributors}}]{SciPy2020}
{Virtanen}, P., {Gommers}, R., {Oliphant}, T.~E., {et~al.} 2020, Nature Methods, 17, 261, \dodoi{10.1038/s41592-019-0686-2}

\bibitem[{{Wooden} {et~al.}(1993){Wooden}, {Rank}, {Bregman}, {Witteborn}, {Tielens}, {Cohen}, {Pinto}, \& {Axelrod}}]{Wooden1993}
{Wooden}, D.~H., {Rank}, D.~M., {Bregman}, J.~D., {et~al.} 1993, \apjs, 88, 477, \dodoi{10.1086/191830}

\bibitem[{{Woosley} {et~al.}(2002){Woosley}, {Heger}, \& {Weaver}}]{Woosley_2002}
{Woosley}, S.~E., {Heger}, A., \& {Weaver}, T.~A. 2002, Reviews of Modern Physics, 74, 1015, \dodoi{10.1103/RevModPhys.74.1015}

\end{thebibliography}
\end{document}